\documentstyle[11pt,epsfig,psfig,amsmath]{article}
\makeatletter

\@addtoreset{equation}{section}

\makeatother

\begin{document}

\title{Casimir energy in the Fulling--Rindler vacuum}
\author{A. A. Saharian$^{1,2}$\footnote{%
Email address: saharyan@server.physdep.r.am} , R. S.
Davtyan$^{1}$,
and A. H. Yeranyan$^{1}$\\
\textit{$^{1}$ Department of Physics, Yerevan State
University, 375049 Yerevan, Armenia,}\\
\textit{$^{2}$ the Abdus Salam International Centre for
Theoretical Physics, 34014 Trieste, Italy }}

\maketitle

\begin{abstract}
The Casimir energy is evaluated for massless scalar fields under
Dirichlet or Neumann boundary conditions, and for the
electromagnetic field with perfect conductor boundary conditions
on one and two infinite parallel plates moving by uniform proper
acceleration through the Fulling--Rindler vacuum in an arbitrary
number of spacetime dimension. For the  geometry of a single plate
both regions of the right Rindler wedge, (i) on the right (RR
region) and (ii) on the left (RL region) of the plate are
considered. The zeta function technique is used, in combination
with contour integral representations. The Casimir energies for
separate RR and RL regions contain pole and finite contributions.
For an infinitely thin plate taking RR and RL regions together, in
odd spatial dimensions the pole parts cancel and the Casimir
energy for the whole Rindler wedge is finite. In $d=3$ spatial
dimensions the total Casimir energy for a single plate is negative
for Dirichlet scalar and positive for Neumann scalar and the
electromagnetic field. The total Casimir energy for two plates
geometry is presented in the form of a sum of the Casimir energies
for separate plates plus an additional interference term. The
latter is negative for all values of the plates separation for
both Dirichlet and Neumann scalars, and for the electromagnetic
field.

\end{abstract}

\bigskip

PACS number(s): 03.70.+k, 11.10.Kk

\bigskip

\section{Introduction}

The Casimir effect is a phenomenon common to all systems
characterized by fluctuating quantities on which external boundary
conditions are imposed. It may have important implications on all
scales, from cosmological to subnuclear. The imposition of
boundary conditions on a quantum field leads to the modification
of the spectrum for the zero--point fluctuations and results in
the shift in the vacuum expectation values for physical quantities
such as the energy density and stresses. In particular, the
confinement of quantum fluctuations causes forces that act on
constraining boundaries. The particular features of the resulting
vacuum forces depend on the nature of the quantum field, the type
of spacetime manifold, the boundary geometries and the specific
boundary conditions imposed on the field. Since the original work
by Casimir in 1948 \cite{Casimir} many theoretical and
experimental works have been done on this problem (see, e.g.,
\cite{Mostepanenko,Plunien,Milton,Lamor,Bordag,Bordag1,Kirs01} and
references therein). Many different approaches have been used:
mode summation method with combination of the zeta function
regularization technique, Green function formalism, multiple
scattering expansions, heat-kernel series, etc. An interesting
topic in the investigations of the Casimir effect is the
dependence of the vacuum characteristics on the type of the
vacuum. It is well known that the uniqueness of the vacuum state is
lost when we work within the framework of quantum field theory in
a general curved spacetime or in non--inertial frames. In
particular, the use of general coordinate transformations in
quantum field theory in flat spacetime leads to an infinite number
of unitary inequivalent representations of the commutation
relations. Different inequivalent representations will in general
give rise to different pictures with different physical
implications, in particular to different vacuum states. For
instance, the vacuum state for an uniformly accelerated observer,
the Fulling--Rindler vacuum \cite{Full73,Full77,Unru76,Boul75},
turns out to be inequivalent to that for an inertial observer, the
familiar Minkowski vacuum (for a mathematical discussion by means of a normal mode analysis see Ref. \cite{Gerl89}). Quantum field theory in accelerated
systems contains many special features produced by a
gravitational field. This fact allows one to avoid some of the difficulties entailed by
renormalization in a curved spacetime. In particular, the near
horizon geometry of most black holes is well approximated by
Rindler and a better understanding of physical effects in this
background could serve as a handle to deal with more complicated
geometries like Schwarzschild. The Rindler geometry shares most of
the qualitative features of black holes and is simple enough to
allow detailed analysis. Another motivation for the investigation
of quantum effects in the Rindler space is related to the fact
that this space is conformally related to the de Sitter space and
to the Robertson--Walker space with negative spatial curvature. As
a result the expectation values of the energy--momentum tensor for
a conformally invariant field and for corresponding conformally
transformed boundaries on the de Sitter and Robertson--Walker
backgrounds can be derived from the corresponding Rindler
counterpart by the standard transformation (see, for instance,
\cite{Birrell}).

The problem of vacuum polarization brought about by the presence
of an infinite plane boundary moving with uniform acceleration
through the Fulling-Rindler vacuum was investigated by Candelas
and Deutsch \cite{Candelas} for the conformally coupled $4D$
Dirichlet and Neumann massless scalar and electromagnetic fields.
In this paper only the region of the right Rindler wedge to the
right of the barrier is considered. In Ref. \cite{Saha02} we have
investigated the Wightman function and the vacuum expectation
values of the energy momentum-tensor for the massive scalar field
with general curvature coupling parameter, satisfying the Robin
boundary conditions on the infinite plane in an arbitrary number
of spacetime dimensions and for the electromagnetic field. Unlike
Ref. \cite{Candelas} we have considered both regions, including
the one between the barrier and Rindler horizon. The vacuum
expectation values of the energy-momentum tensors for scalar and
electromagnetic fields for the geometry of two parallel plates
moving by uniform acceleration are investigated in Ref.
\cite{Avag02}. In particular, the vacuum forces acting on the
boundaries are evaluated. They are presented as a sum of the
interaction and self-action parts. The interaction forces between
the plates are always attractive for both scalar and
electromagnetic cases. The self-action forces contain well-known
surface divergencies and needs further regularization. In Refs.
\cite{Saha02,Avag02} the mode summation method is used in
combination with the generalized Abel-Plana summation formula
\cite{Sahrev}. This allowed us to present the vacuum expectation
values in the terms of the purely Rindler and boundary parts. Due
to the well known non-integrable surface divergences in the
boundary parts, the total Casimir energy cannot be obtained by
direct integration of the vacuum energy density and needs an
additional regularization. Many regularization techniques are
available nowadays and, depending on the specific physical problem
under consideration, one of them may be more suitable than the
others. In particular, the generalized zeta function method
\cite{Dowk76,Eliz94} is in general very powerful to give physical
meaning to the divergent quantities. There are several examples of
the application of this method to the evaluation of the Casimir
effect (see, for instance,
\cite{Bordag1,Eliz94,Blau88,Eliz93,Lese94,Rome95,Bord96,Lese96,Bord96b,Bord97,
Lamb99,Cogn01}). In this paper, by using the zeta function
technique, we consider the Casimir energy for the geometries of a
single and two parallel plates, moving by uniform proper
acceleration through the Fulling-Rindler vacuum.

The paper is organized as follows.  In Sec. \ref{sec:Dir1pl} the
Casimir energy is evaluated for the Dirichlet scalar in the case
of a single plate. For the both RR and RL regions we construct
integral representations of the related zeta functions and
analytically continue them to the physical region. The geometry of
two plates in the case of the Dirichlet scalar is investigated in
Sec. \ref{sec:Dir2pl}. We show that the corresponding zeta
function is a sum of the zeta functions for the separate plates
plus an additional interference term which is finite in the
physical region. Similar problems for the Neumann scalar are
investigated in Sec. \ref{sec:neuscal} and Sec. \ref{sec:Neu2pl},
respectively. Section \ref{sec:elmag} considers the Casimir energy
for the electromagnetic field assuming that the plates are perfect
conductors. Section \ref{sec:conc} concludes the main results of the
paper. In Appendix \ref{ap:d1} the case $d=1$ is considered
separately. Appendix \ref{ap:propE} discusses the relation between
the calculated energies and the energies measured by an uniformly
accelerated observer.

\section{Casimir energy for a single Dirichlet plate}\label{sec:Dir1pl}

Consider a real massless scalar field $\varphi (x)$ with curvature
coupling parameter $\zeta $ satisfying the field equation
\begin{equation}
\nabla _{\mu }\nabla ^{\mu }\varphi +\zeta R\varphi =0,
\label{fieldeq}
\end{equation}
with $R$ being the scalar curvature for a $d+1$--dimensional
background spacetime, $\nabla _{\mu }$ is the covariant derivative
operator associated with the corresponding metric tensor $g_{\mu
\nu }$. For minimally and conformally coupled scalars one has
$\zeta =0$ and $\zeta =(d-1)/4d$, respectively. Our main interest
in this paper will be the Casimir energy in the Rindler spacetime
induced by a single and two parallel plates moving with uniform
proper acceleration when the quantum field is prepared in the
Fulling-Rindler vacuum (note the difference of our problem from that related to the Unruh effect, where the field is in its Minkowski vacuum state). For this problem the background spacetime
is flat and in Eq. (\ref{fieldeq}) we have $R=0$. As a result the
eigenmodes are independent of the curvature coupling parameter
and, hence, the Casimir energy will not depend on this parameter.
However, the local characteristics of the vacuum such as energy
density and vacuum stresses depend on the parameter $\zeta $
\cite{Saha02,Avag02}.

In the accelerated frame it is convenient to introduce Rindler coordinates $%
(\tau ,\xi ,{\bf x})$ which are related to the Minkowski ones,
$(t,x^{1},{\bf x})$ by transformations
\begin{equation}
t=\xi \sinh \tau ,\quad x^{1}=\xi \cosh \tau ,  \label{RindMin}
\end{equation}
where ${\bf x}=(x^{2},\ldots ,x^{d})$ denotes the set of
coordinates parallel to the plate. In these coordinates the
Minkowski line element takes the form
\begin{equation}
ds^{2}=\xi ^{2}d\tau ^{2}-d\xi ^{2}-d{\bf x}^{2},  \label{metric}
\end{equation}
and a wordline defined by $\xi ,{\bf x}={\rm const}$ describes an
observer with constant proper acceleration $\xi ^{-1}$. Rindler
time coordinate $\tau $ is proportional to the proper time along a
family of uniformly accelerated trajectories which fill the
Rindler wedge, with the proportionality constant equal to the
acceleration. Let $\{\varphi _{\alpha }(x),\varphi _{\alpha
}^{\ast }(x)\}$ be a complete set of positive and negative
frequency solutions to the field equation (\ref{fieldeq}), where
$\alpha $ denotes a set of quantum numbers.
\begin{figure}[tbph]
\begin{center}
\epsfig{figure=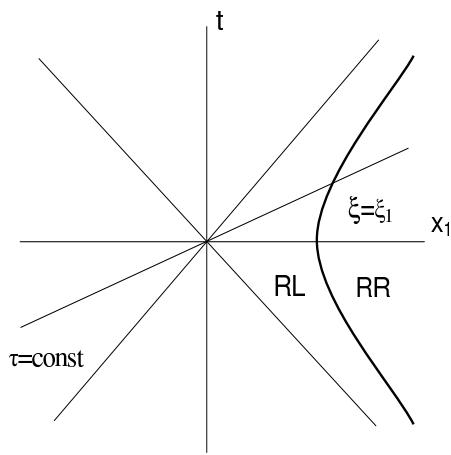,width=6cm,height=6cm}
\end{center}
\caption{The $(x_{1},t)$ plane with the Rindler coordinates. The
heavy line $\protect\xi =\protect\xi _{1}$ represent the
trajectory of the plate.} \label{fig11pl}
\end{figure}
For the geometry under consideration the metric and boundary
conditions are static and translational invariant in the
hyperplane parallel to the plates. It follows from here that the
corresponding part of the eigenfunctions can be taken in the
standard plane wave form:
\begin{equation}
\varphi _{\alpha }=C\phi (\xi )\exp \left[ i\left( {\bf kx}-\omega
\tau \right) \right] ,\quad \alpha =({\bf k},\omega ),\quad {\bf
k}=(k_{2},\ldots ,k_{d}).  \label{wavesracture}
\end{equation}
The equation for $\phi (\xi )$ is obtained from field equation (\ref{fieldeq}%
) on background of metric (\ref{metric}):
\begin{equation}
\xi ^{2}\phi ^{\prime \prime }(\xi )+\xi \phi ^{\prime }(\xi
)+\left( \omega ^{2}-k^{2}\xi ^{2}\right) \phi (\xi )=0,
\label{fiequ}
\end{equation}
where the prime denotes a differentiation with respect to the argument, and $%
k=|{\bf k}|$. The linearly independent solutions to equation
(\ref{fiequ}) are the Bessel modified functions $I_{i\omega }(k\xi
)$ and $K_{i\omega }(k\xi )$ of the imaginary order. The
eigenfrequencies are determined from the boundary conditions
imposed on the field on the bounding surfaces.

\subsection{Vacuum energy in the RR region}\label{subsec:dirRR}

In this section we will consider the vacuum energy for a scalar
field satisfying Dirichlet boundary condition on a single plate
located at $\xi =\xi _{1}$:
\begin{equation}
\varphi |_{\xi =\xi _{1}}=0.  \label{DirR1pl}
\end{equation}
We will assume that the plate is situated in the right Rindler
wedge $x^{1}>\left| t\right| $. The surface $\xi =\xi _{1}$
represents the trajectory of the boundary, which therefore has
proper accelerations $\xi _{1}^{-1}$ (see Fig. \ref{fig11pl}).
This trajectory divides the right Rindler wedge into two regions
with $\xi \geq \xi _1$ and $\xi \leq \xi _1$. In the following we
will refer these regions as RR and RL regions, respectively. First
let us consider the vacuum energy in the RR region. In this
region, for a complete set of solutions that are of positive
frequency with respect to $\partial /\partial \tau $ and bounded
as $\xi \rightarrow \infty $, in Eq. (\ref{wavesracture}) one has to take $%
\phi (\xi )=K_{i\omega }(k\xi )$. For the Dirichlet scalar the
corresponding eigenfrequencies are determined from the equation
\begin{equation}
K_{i\omega }(k\xi _{1})=0.  \label{Kiom}
\end{equation}
The positive roots to this equation arranged in the ascending
order we will denote by $\omega =\omega _{1Dn}(k\xi _{1})$,
$n=1,2,\ldots $, $\omega _{1Dn}<\omega _{1Dn+1}$. To distinguish
in the notations we will use indices $D$ and $N$ to denote
Dirichlet and Neumann boundary conditions respectively. Similarly
the quantities for the RR and RL regions will be denoted by
indices $R$ and $L$, respectively. The vacuum energy per unit
surface of the plate in the region $\xi \geq \xi _{1}$ is given by
formula
\begin{equation}
E_{1D}^{(R)}(\xi _{1})=\frac{1}{2}\int \frac{d^{d-1}k}{(2\pi )^{d-1}}%
\sum_{n=1}^{\infty }\omega _{1Dn}(k\xi _{1})=\frac{B_d}{\xi
_{1}^{d-1}}\int_{0}^{\infty }dx\,x^{d-2}\sum_{n=1}^{\infty }\omega
_{1Dn}(x),  \label{E1DR}
\end{equation}
where we have integrated over angular coordinates, and
\begin{equation}\label{Bd}
B_d=\frac{1}{(4\pi )^{\frac{d-1}{2}}\Gamma \left(
\frac{d-1}{2}\right) } .
\end{equation}
Here and below we will assume that $d>1$. The case $d=1$ is
considered in Appendix \ref{ap:d1}. As it stands, the right-hand
side of equation (\ref{E1DR}) clearly diverges and needs some
regularization. We regularize it by defining the function
\begin{equation}
Z_{DK}(s)=\int_{0}^{\infty }dx\,x^{d-2}\zeta _{1D}^{(R)}(s,x)
\label{Zs0}
\end{equation}
where we have introduced the partial zeta function related to the
eigenfrequences defined by Eq. (\ref{Kiom}):
\begin{equation}
\zeta _{1D}^{(R)}(s,x)=\sum_{n=1}^{\infty }\omega _{1Dn}^{-s}(x).
\label{zet1D}
\end{equation}
The Casimir energy is expressed as
\begin{equation}
E_{1D}^{(R)}(\xi _{1})=\frac{B_d}{\xi
_{1}^{d-1}}Z_{DK}(s)|_{s=-1}. \label{EnZs}
\end{equation}
In accordance with this formula, the computation of the Casimir
energy requires the analytic continuation of the zeta function to
the value $s=-1$.

The starting point of our consideration is the representation of
the partial zeta function for the corresponding modes in term of
contour integral (for a similar treatment of the zeta function as
a contour integral see Refs.
\cite{Eliz93,Lese94,Rome95,Bord96,Lese96,Bord96b,Bord97,Lamb99,Cogn01,Kame92,
Barv92}):
\begin{equation}
\zeta _{1D}^{(R)}(s,x)=\frac{1}{2\pi i}\int_{C}dz\,z^{-s}\frac{\partial }{%
\partial z}\ln K_{iz}(x),  \label{zet1D1}
\end{equation}
where $C$ is a closed counterclockwise contour in the complex $z$
plane enclosing all zeros $\omega _{1Dn}(x)$. We assume that this
contour is made of a large semicircle (with radius tending to
infinity) centered at the origin and placed to its right, plus a
straight part overlapping the imaginary axis. When the radius of
the semicircle tends to infinity the corresponding contribution
into $\zeta _{1D}^{(R)}(s,x)$ vanishes for ${\mathrm{ Re}}\, s>1$.
After parametrizing the integrals over imaginary axis we arrive at
the expression
\begin{equation}
\zeta _{1D}^{(R)}(s,x)=\frac{1}{\pi }\sin \frac{\pi
s}{2}\int_{0}^{\infty }dz\,z^{-s}\frac{\partial }{\partial z}\ln
K_{z}(x).  \label{zet1D2}
\end{equation}
For small $z$, $z\rightarrow 0$, one has $K_{z}(x)\approx
K_{0}(x)+(\partial ^{2}K_{z}(x)/\partial z^{2})_{z=0}(z^{2}/2)$,
and the integral in (\ref{zet1D2}) converges at the lower limit
for ${\mathrm{Re}}\, s<2$. Hence, this integral representation of the
zeta function is valid for $1<{\mathrm{Re}}\, s<2$. To evaluate the
Casimir energy we need the zeta function at $s=-1$ and so we have
to do an analytic continuation of Eq. (\ref{zet1D2}). To this aim
we employ the uniform asymptotic expansion of the modified Bessel
function $K_z(x)$ for large values of the order \cite{Abramowitz}:
\begin{equation}
K_{z}(x)=\sqrt{\frac{\pi }{2}}\frac{e^{-z\eta (x/z)}}{\left(
x^{2}+z^{2}\right) ^{1/4}}K_{z}^{(D)}(x),\quad K_{z}^{(D)}(x)\sim
\sum_{l=0}^{\infty }(-1)^{l}\frac{\overline{u}_{l}(t)}{\left(
x^{2}+z^{2}\right) ^{l/2}},  \label{Kzas}
\end{equation}
where
\begin{equation}
t=\frac{z}{\sqrt{x^{2}+z^{2}}},\quad \eta (x)=\sqrt{1+x^{2}}+\ln \frac{x}{1+%
\sqrt{1+x^{2}}},\quad \overline{u}_{l}(t)=\frac{u_{l}(t)}{t^{l}},
\label{tu}
\end{equation}
and the expressions for the functions $u_{l}(t)$ are given in
\cite{Abramowitz}. From these expressions it follows that the
coefficients $ \overline{u}_{l}(t)$ have the structure
\begin{equation}
\overline{u}_{l}(t)=\sum_{m=0}^{l}u_{lm}t^{2m},  \label{ulbar}
\end{equation}
with the numerical coefficients $u_{lm}$. From the recurrence
relations for the polynomials $u_{l}(t)$ (see \cite{Abramowitz}),
the following recurrence formulae can be obtained for the
coefficients $u_{lm}$:
\begin{equation}
u_{l+1,m}=\frac{1}{2}u_{lm}\left[ 2m+l+\frac{1}{4(2m+l+1)}\right] -\frac{1}{2%
}u_{l,m-1}\left[ 2m+l-2+\frac{5}{4(2m+l+1)}\right] , \label{recbl}
\end{equation}
where $m=0,1,\ldots ,l+1$, and $u_{l,-1}=u_{l,l+1}=0$, $u_{00}=1$.
Now making use the expansion (\ref{Kzas}), the zeta function can
be presented in the form
\begin{equation}
\zeta _{1D}^{(R)}(s,x)=\zeta _{1D}^{(R0)}(s,x)+\zeta
_{1D}^{(R1)}(s,x), \label{zet1D3}
\end{equation}
where
\begin{eqnarray}
\zeta _{1D}^{(R0)}(s,x) &=&\frac{1}{\pi }\sin \frac{\pi s}{2}%
\int_{0}^{\infty }dz\,z^{-s}\frac{\partial }{\partial z}\ln
\sqrt{\frac{\pi }{2}}\frac{e^{-z\eta (x/z)}}{\left(
x^{2}+z^{2}\right) ^{1/4}},
\label{zet1D01} \\
\zeta _{1D}^{(R1)}(s,x) &=&\frac{1}{\pi }\sin \frac{\pi s}{2}%
\int_{0}^{\infty }dz\,z^{-s}\frac{\partial }{\partial z}\ln
K_{z}^{(D)}(x). \label{zet1D11}
\end{eqnarray}
Under the conditions $1<{\mathrm{Re}}s<2$, the term (\ref
{zet1D01}) can be easily evaluated to give
\begin{equation}
\zeta _{1D}^{(R0)}(s,x)=\frac{(x/2)^{1-s}}{\pi (1-s)}B\left( 1-s,\frac{s-1}{2}%
\right) \sin \frac{\pi s}{2}-\frac{x^{-s}}{4},  \label{zet1D02}
\end{equation}
with the beta function $B(x,y)$. Now using the standard
dimensional regularization result that the renormalized value of
the integrals of the type $\int_{0}^{\infty }dx\,x^{\beta }$ is
equal to zero (see, e.g., \cite {Coll84}), we conclude that the
contribution of the term $\zeta _{1D}^{(R0)}(s,x)$ into Eq.
(\ref{Zs0}) vanishes. This can be seen by another way, considering the case of a scalar field with nonzero mass $m$ and taking the limit $m\to 0$ after the evaluation of the corresponding integrals (for this trick in the calculations of the Casimir energy see, for instance, Refs. \cite{Bord02,Nest03}). For the massive case in Eq. (\ref{wavesracture}) one has $\phi (\xi )=K_{i\omega }(\xi \sqrt{k^2+m^2} )$ and the corresponding formulae for $\zeta _{1D}^{(R)}(s,x)$ are obtained from those given above in this section by replacement $x\to \sqrt{x^2+m^2\xi _1^2}$. With this replacement the integral corresponding to the contribution of $\zeta _{1D}^{(R0)}$ into Eq. (\ref{Zs0}) can be easily evaluated in terms of gamma function and vanishes in the limit $m\to 0$ for ${\mathrm{Re}}\, s<d-1$. For this reason in the following we will
concentrate on the contribution of the second term on the right of
Eq. (\ref{zet1D3}). The corresponding expression for $Z_{DK}(s)$
from Eq. (\ref{Zs0}) takes the form
\begin{equation}
Z_{DK}(s)=\frac{1}{\pi }\sin \frac{\pi s}{2}\int_{0}^{\infty
}dx\,x^{d-2}\int_{0}^{\infty }dz\,z^{-s}\frac{\partial }{\partial
z}\ln K_{z}^{(D)}(x).  \label{Zs}
\end{equation}
To deal with one infinite limit integral in the following it will
be convenient to introduce polar coordinates on the plane $(z,x)$:
\begin{equation}
z=r\cos \theta ,\quad x=r\sin \theta ,  \label{rtet}
\end{equation}
In these coordinates one has the following expressions
\begin{equation}
t=\cos \theta ,\quad z\eta (x/z)=rg(\theta ) ,\qquad
K_{z}^{(D)}(x)\sim \sum_{l=0}^{\infty
}(-1)^{l}\frac{\overline{u}_{l}(\cos \theta )}{r^{l}}.
\label{tpol}
\end{equation}
where we have introduced the notation
\begin{equation}
g(\theta )=1+\cos \theta \,\ln \frac{\sin \theta }{1+\cos \theta
}. \label{gtet}
\end{equation}
The integral representation (\ref{Zs}) is well suited for the
analytic continuation in $s$. Following the usual procedure
applied in the analogous calculations (see, for instance,
\cite{Bordag1} and references therein), we subtract and add to the
integrand in Eq. (\ref{Zs}) $M$ leading terms of the corresponding
asymptotic expansion and exactly integrate the asymptotic part.
For this let us define the coefficients $U_{l}(\cos \theta )$ in
accordance with
\begin{equation}
\ln \left( \sum_{l=0}^{\infty }(-1)^{l}\frac{\overline{u}_{l}(\cos \theta )}{%
r^{l}}\right) =\sum_{l=1}^{\infty }(-1)^{l}\frac{U_{l}(\cos \theta
)}{\left( 1+r^{2}\right) ^{l/2}}.  \label{defUl}
\end{equation}
Here on the right we have expanded over $\left( 1+r^{2}\right)
^{-1/2}$ to avoid convergence problems in the integral over $r$ at
the lower limit $r=0$. Hence, Eq. (\ref{Zs}) may be split into the
following pieces
\begin{equation}
Z_{DK}(s)=Z_{DK}^{(as)}(s)+Z_{DK}^{(1)}(s),  \label{Zs1}
\end{equation}
with
\begin{eqnarray}
Z_{DK}^{(as)}(s) &=&\frac{1}{\pi }\sin \frac{\pi
s}{2}\int_{0}^{\infty
}dx\,x^{d-2}\int_{0}^{\infty }dz\,z^{-s}\frac{\partial }{\partial z}%
\sum_{l=1}^{M}(-1)^{l}\frac{U_{l}(\cos \theta )}{\left( 1+r^{2}\right) ^{l/2}%
},  \label{Zsas1} \\
Z_{DK}^{(1)}(s) &=&\frac{1}{\pi }\sin \frac{\pi
s}{2}\int_{0}^{\infty }dx\,x^{d-2}\int_{0}^{\infty
}dz\,z^{-s}\frac{\partial }{\partial z}\left( \ln
K_{z}^{(D)}(x)-\sum_{l=1}^{M}(-1)^{l}\frac{U_{l}(\cos \theta
)}{\left( 1+r^{2}\right) ^{l/2}}\right) .  \label{ZDK1Ns}
\end{eqnarray}
The number of terms $M$ in these formulae is determined by the spatial
dimension of the problem under consideration. For $M\geq d$ the
expression $Z_{DK}^{(1)}(s)$ is finite at $s=-1$ and, hence, for
our aim it is sufficient to subtract $M=d$ asymptotic terms. For
larger $M$ the convergence is stronger. This means that with
$M\geq d$ we can directly put $s=-1$ in the expression for
$Z_{DK}^{(1)}(s)$ in Eq. (\ref{ZDK1Ns}) and perform the integral
numerically. For numerical evaluation it is useful to
integrate by parts the $z$-integral and then to introduce polar coordinates
in the $(z,x)$-plane. This yields
\begin{equation}
Z_{DK}^{(1)}(-1)=\frac{1}{\pi }\int_{0}^{\infty
}dr\,r^{d-1}\int_{0}^{\pi
/2}d\theta \,\sin ^{d-2}\theta \left[ \ln \left( \sqrt{\frac{2r}{\pi }}%
e^{rg(\theta )}K_{r\cos \theta }(r\sin \theta )\right)
-\sum_{l=1}^{M}(-1)^{l}\frac{U_{l}(\cos \theta )}{\left(
1+r^{2}\right) ^{l/2}}\right] .  \label{ZDK1-1}
\end{equation}

Now we turn to the asymptotic part of \ $Z_{DK}(s)$ given by
expression (\ref{Zsas1}). For this term the analytic continuation
can be done analytically because its structure is simple. To
evaluate this part we integrate over $z$ by parts and note that
the functions $U_{l}(t)$ are polynomials in $t$:
\begin{equation}
U_{l}(t)=\sum_{m=0}^{l}U_{lm}t^{2m},  \label{Ul}
\end{equation}
where the numerical coefficients $U_{lm}$ are related to the
coefficients in Eq. (\ref{ulbar}) by expansion (\ref{defUl}). The
first five functions are
\begin{subequations}\label{U123}
\begin{eqnarray}
U_{1}(t) &=&\frac{1}{8}-\frac{5}{24}t^{2},\quad U_{2}(t)=\frac{1}{16}-\frac{3%
}{8}t^{2}+\frac{5}{16}t^{4},   \\
U_{3}(t) &=&\frac{49}{384}-\frac{1793}{1920}t^{2}+\frac{221}{128}t^{4}-\frac{%
1105}{1152}t^{6},  \\
U_{4}(t) &=&\frac{21}{128}-\frac{83}{32}t^{2}+\frac{551}{64}t^{4}-\frac{339}{%
32}t^{6}+\frac{565}{128}t^{8}, \\
U_{5}(t) &=&\frac{1813}{5120}-\frac{297649}{35840}t^{2}+\frac{198755}{4608}%
t^{4}-\frac{136907}{1536}t^{6}+\frac{82825}{1024}t^{8}-\frac{82825}{3072}%
t^{10}.
\end{eqnarray}
\end{subequations}
Taking into account Eq. (\ref{Ul}) and introducing polar
coordinates, for the asymptotic part one finds
\begin{equation}
Z_{DK}^{(as)}(s)=\frac{s}{\pi }\sin \frac{\pi s}{2}\sum_{l=1}^{M}(-1)^{l}%
\sum_{m=0}^{l}U_{lm}\int_{0}^{\infty }dr\frac{r^{d-s-2}}{\left(
1+r^{2}\right) ^{l/2}}\int_{0}^{\pi /2}d\theta \sin ^{d-2}\theta
\cos ^{2m-s-1}\theta .  \label{Zsas2}
\end{equation}
Evaluating the integrals by using the standard formulae (see, for
instance, \cite{Prud}), we arrive at the expression
\begin{equation}
Z_{DK}^{(as)}(s)=\frac{s}{4\pi }\sin \frac{\pi s}{2}\sum_{l=1}^{M}(-1)^{l}%
\sum_{m=0}^{l}U_{lm}B\left(
\frac{d-s-1}{2},\frac{l+s-d+1}{2}\right) B\left(
m-\frac{s}{2},\frac{d-1}{2}\right) \ ,  \label{Zsas3}
\end{equation}
where the pole contributions are given explicitly in terms of beta
function. In the sum over $l$, the terms with even $d-l\geq 0$
have simple poles at $s=-1$ coming from the first beta function.
Introducing a new summation variable $p=(d-l)/2$, the
corresponding residue can be easily found by using the standard
formula for the gamma function:
\begin{equation}
Z_{DK,-1}^{(as)}=\frac{(-1)^{d}}{\pi d}\sum_{p=0}^{p_{d}}(-1)^{p}%
\sum_{m=0}^{d-2p}U_{d-2p,m}\frac{B\left(
m+\frac{1}{2},\frac{d-1}{2}\right) }{B(\frac{d}{2}-p,p+1)} \ ,
\label{Zsres}
\end{equation}
where
\begin{equation}
p_{d}=\left\{
\begin{array}{cc}
(d-1)/2, & \quad {\rm for\;odd}\;d \\
d/2-1 & \quad {\rm for\;even}\;d
\end{array}
\right. .  \label{pd}
\end{equation}
Hence, Laurent-expanding near $s=-1$ we can write
\begin{equation}
Z_{DK}^{(as)}(s)=\frac{Z_{DK,-1}^{(as)}}{s+1}+Z_{DK,0}^{(as)}+O(s+1),
\label{Zsas4}
\end{equation}
with
\begin{eqnarray}
Z_{DK,0}^{(as)} &=&\frac{(-1)^{d}}{2\pi d}\sum_{p=0}^{p_{d}}(-1)^{p}%
\sum_{m=0}^{d-2p}U_{d-2p,m}\frac{B\left(
m+\frac{1}{2},\frac{d-1}{2}\right)
}{B(\frac{d}{2}-p,p+1)}\nonumber \\
& & \times \left[ \psi (p+1)-\psi \left( \frac{d}{2}\right) +\psi
\left( m+\frac{d}{2}\right) -\psi \left( m+\frac{1}{2}\right)
-2\right] \nonumber
\\
&&+\frac{1}{4\pi }\left( \sum_{l=1,d-l={\rm odd}}^{d-1}+\sum_{l=d+1}^{M}%
\right) (-1)^{l}\sum_{m=0}^{l}U_{lm}B\left(
\frac{d}{2},\frac{l-d}{2}\right) B\left(
m+\frac{1}{2},\frac{d-1}{2}\right) \label{ZDKas0} \ ,
\end{eqnarray}
where $\psi (x)=d\ln \Gamma (x)/dx$ is the diagamma function and
the second sum in the braces of the third line is present only for
$M\geq d+1$. The first term on the right of Eq. (\ref{ZDKas0})
with diagamma functions comes from the finite part of the Laurent
expansion of the summands with even $d-l\geq 0$ in Eq.
(\ref{Zsas3}). Gathering all contributions together, near $s=-1$
we find
\begin{equation}
Z_{DK}(s)=\frac{Z_{DK,-1}^{(as)}}{s+1}%
+Z_{DK,0}^{(as)}+Z_{DK}^{(1)}(-1)+O(s+1),  \label{ZDKexp}
\end{equation}
where the separate terms are defined by formulae (\ref{ZDK1-1}), (\ref{Zsres}%
), (\ref{ZDKas0}). Using this result, for the vacuum energy
induced by a single plate at $\xi =\xi _1$ in the region $\xi \geq
\xi _1$ one receives
\begin{equation}\label{E1DR3}
    E_{1D}^{(R)}(\xi _1)=E_{1Dp}^{(R)}+E_{1Df}^{(R)},
\end{equation}
where for the pole and finite contributions one has
\begin{equation}\label{E1DRp}
E_{1Dp}^{(R)}= \frac{B_dZ_{DK,-1}^{(as)}}{\xi _1^{d-1} (s+1)},
\quad E_{1Df}^{(R)}= \frac{B_d}{\xi _1^{d-1} }
\left[Z_{DK,0}^{(as)}+Z_{DK}^{(1)}(-1)\right] .
\end{equation}
The numerical results corresponding to the vacuum energy
(\ref{E1DR3}) with separate parts (\ref{E1DRp}) are presented in
the first two column of Table \ref{tab:diren} for spatial
dimensions $d=2,3,4$.
\begin{table}
  \centering
  \caption{\label{tab:diren} Pole and finite parts of the Casimir energy for a single
  Dirichlet plate.}

  \begin{tabular}{cccccc}
  \hline
  \hline
  $d$ & $\xi _1^{d-1}E_{1Dp}^{(R)}$ & $\xi _1^{d-1}E_{1Df}^{(R)}$ &
  $\xi _1^{d-1}E_{1Dp}^{(L)}$ & $\xi _1^{d-1}E_{1Df}^{(L)}$&
  $\xi _1^{d-1}E_{1Df}$ \\
  \hline
  2 & $-\frac{1}{512\pi (s+1)}$ & 0.00185 & $-\frac{1}{512\pi (s+1)}$ & -0.00269 & -0.000836 \\

  3 & $\frac{1}{1260\pi ^2(s+1)}$ & -0.000620 & $-\frac{1}{1260\pi ^2(s+1)} $& -0.000695 &
  -0.00131  \\

  4 & $-\frac{37}{262144\pi ^2 (s+1)}$ & 0.000225 & $-\frac{37}{262144\pi ^2 (s+1)}$ &
  -0.000235 & -0.00000996  \\
  \hline
  \hline
\end{tabular}

\end{table}

In Appendix \ref{ap:propE} we have discussed the relation between
vacuum energy (\ref{E1DR3}) and the energy measured by an
uniformly accelerated observer with the proper acceleration $g$.
These energies are connected by formula (\ref{EgErel}), where $\mu
$ is an arbitrary constant with the dimension of mass. Using the
numerical results given in Table \ref{tab:diren}, the Casimir
energy in the RR region for the massless Dirichlet scalar in $d=3$
measured by an uniformly accelerated observer is presented in the
form
\begin{equation}\label{ER1Dg}
E^{(R)}_{1Dg}(\xi _1)=\frac{g}{\xi _1^2} \left(
-0.000620+\frac{1}{1260\pi ^2}\left[ \frac{1}{s+1}+\ln \left(
\frac{\mu }{g}\right) \right] \right) ,
\end{equation}
where the logarithmic term is a consequence of the divergence and
has to be viewed as a remainder of the renormalization process.
The discussion for the role of the normalization scale $\mu $ in
the calculations of the Casimir energy can be found in Ref.
\cite{Blau88}. The remained pole term in the Casimir energy is a
characteristic feature for the zeta function regularization method
and has been found for many cases of boundary geometries. Note
that a very powerful tool for studying divergence structure of the
vacuum energy is the heat kernel expansion (see, for instance,
\cite{Bordag1,Vass03} and references therein). In particular, the
coefficient of the logarithmic term is determined by the
corresponding boundary coefficient in the heat kernel asymptotic
expansion.

\subsection{Vacuum energy in the RL region}\label{subsec:dirRL}

Now we turn to the Casimir energy in the region between a single
plate and the Rindler horizon corresponding to $\xi =0$. As in
previous subsection we will assume that the plate is located at
$\xi =\xi _{1}$ and the field satisfies Dirichlet boundary
condition (\ref{DirR1pl}) on it. In the next section we show that
the corresponding vacuum energy can be determined by the formula
\begin{equation}
E_{1D}^{(L)}(\xi _{1})=\frac{B_d}{\xi _1^{d-1} }Z_{DI}(s)|_{s=-1},
\label{EL1D}
\end{equation}
where we have defined the function
\begin{equation}
Z_{DI}(s)=\int_{0}^{\infty }dx\,x^{d-2}\zeta _{1D}^{(L)}(s,x),
\label{ZDlI}
\end{equation}
with the partial zeta function
\begin{equation}
\zeta _{1D}^{(L)}(s,x)=\frac{1}{\pi }\sin \frac{\pi
s}{2}\int_{\rho }^{\infty }dz\,z^{-s}\frac{\partial }{\partial
z}\ln I_{z}(x). \label{zetaLD}
\end{equation}
Here $\rho $ is a small number which we will put zero later.
Unlike to the case (\ref{zet1D2}), here we can not directly put
$\rho =0$, as for values $s $ with convergence at the upper limit,
the $z$-integral will diverge at the lower limit. The analytical
continuation for (\ref{zetaLD}) as a function on $s$ can be done
by the way similar to that given above for the region on the right
of a single plate. First, we note that for the Bessel modified
function one has the uniform asymptotic expansion
\cite{Abramowitz}
\begin{equation}
I_{z}(x)=\frac{1}{\sqrt{2\pi }}\frac{e^{z\eta (x/z)}}{\left(
x^{2}+z^{2}\right) ^{1/4}}I_{z}^{(D)}(x),\quad I_{z}^{(D)}(x)\sim
\sum_{l=0}^{\infty }\frac{\overline{u}_{l}(t)}{\left(
x^{2}+z^{2}\right) ^{l/2}},  \label{Izas}
\end{equation}
where the functions $\eta (x)$, $t$, $\overline{u}_{l}(t)$ are
defined in Eq. (\ref{tu}). This allows us to present the partial
zeta function in the form
\begin{equation}
\zeta _{1D}^{(L)}(s,x)=\zeta _{1D}^{(L0)}(s,x)+\zeta
_{1D}^{(L1)}(s,x), \label{zet1LD}
\end{equation}
with
\begin{eqnarray}
\zeta _{1D}^{(L0)}(s,x) &=&\frac{1}{\pi }\sin \frac{\pi
s}{2}\int_{\rho }^{\infty }dz\,z^{-s}\frac{\partial }{\partial
z}\ln \frac{e^{z\eta (x/z)}}{\sqrt{2\pi }\left( x^{2}+z^{2}\right)
^{1/4}}, \label{zetL1D01}
\\
\zeta _{1D}^{(L1)}(s,x) &=&\frac{1}{\pi }\sin \frac{\pi
s}{2}\int_{\rho }^{\infty }dz\,z^{-s}\frac{\partial }{\partial
z}\ln I_{z}^{(D)}(x).\label{zetL1D11}
\end{eqnarray}
For $1<{\mathrm{Re}}s<2$ the $z$-integral in the expression for
$\zeta _{1D}^{(L0)}(s,x)$ converges in the limit $\rho \rightarrow
0$ and in this limit the evaluation of the integral gives
\begin{equation}
\zeta _{1D}^{(R0)}(s,x)=-\frac{(x/2)^{1-s}}{\pi (1-s)}B\left( 1-s,\frac{s-1}{2}%
\right) \sin \frac{\pi s}{2}-\frac{x^{-s}}{4}.  \label{zetL1D02}
\end{equation}
As in the case of the RR region the contribution of this term into
the vacuum energy vanishes.

Now we consider the part due to Eq. (\ref{zetL1D11}):
\begin{equation}
Z_{DI}(s)=\frac{1}{\pi }\sin \frac{\pi s}{2}\int_{0}^{\infty
}dx\,x^{d-2}\int_{\rho }^{\infty }dz\,z^{-s}\frac{\partial
}{\partial z}\ln I_{z}^{(D)}(x).  \label{ZDI}
\end{equation}
By using asymptotic expansion (\ref{Izas}), we can present it in
the form
\begin{equation}
Z_{DI}(s)=Z_{DI}^{(as)}(s)+Z_{DI}^{(1)}(s),  \label{ZDI1}
\end{equation}
where
\begin{eqnarray}
Z_{DI}^{(as)}(s) &=&\frac{1}{\pi }\sin \frac{\pi
s}{2}\int_{0}^{\infty
}dx\,x^{d-2}\int_{\rho }^{\infty }dz\,z^{-s}\frac{\partial }{\partial z}%
\sum_{l=1}^{M}\frac{U_{l}(\cos \theta )}{\left( 1+r^{2}\right)
^{l/2}},
\label{ZDI1as} \\
Z_{DI}^{(1)}(s) &=&\frac{1}{\pi }\sin \frac{\pi
s}{2}\int_{0}^{\infty
}dx\,x^{d-2}\int_{\rho }^{\infty }dz\,z^{-s}\frac{\partial }{\partial z}%
\left( \ln I_{z}^{(D)}(x)-\sum_{l=1}^{M}\frac{U_{l}(\cos \theta
)}{\left( 1+r^{2}\right) ^{l/2}}\right) .  \label{ZDI11}
\end{eqnarray}
In the expression for $Z_{DI}^{(as)}(s)$ we can directly put $\rho
=0$, and after introducing polar coordinates and integrating, one
finds
\begin{equation}
Z_{DI}^{(as)}(s)=\frac{s}{4\pi }\sin \frac{\pi s}{2}\sum_{l=1}^{M}%
\sum_{m=0}^{l}U_{lm}B\left(
\frac{d-s-1}{2},\frac{l+s-d+1}{2}\right) B\left(
m-\frac{s}{2},\frac{d-1}{2}\right) .  \label{ZDIas2}
\end{equation}
At $s=-1$ this function has a simple pole with the residue
\begin{equation}
Z_{DI,-1}^{(as)}=(-1)^{d} Z_{DK,-1}^{(as)},  \label{ZDIres}
\end{equation}
where $Z_{DK,-1}^{(as)}$ is determined by formula (\ref{Zsres}).
Now we have the following Laurent expansion
\begin{equation}
Z_{DI}^{(as)}(s)=\frac{Z_{DI,-1}^{(as)}}{s+1}+Z_{DI,0}^{(as)}+O(s+1),
\label{ZDIexp}
\end{equation}
where
\begin{eqnarray}
Z_{DI,0}^{(as)} &=&\frac{1}{2\pi d}\sum_{p=0}^{p_{d}}(-1)^{p}%
\sum_{m=0}^{d-2p}U_{d-2p,m}\frac{B\left(
m+\frac{1}{2},\frac{d-1}{2}\right) }{B(\frac{d}{2}-p,p+1)}
\nonumber \\
&& \times \left[ \psi (p+1)-\psi \left( \frac{d}{2}\right) +\psi
\left( m+\frac{d}{2}\right) -\psi \left( m+\frac{1}{2}\right)
-2\right]
\nonumber \\
&&+\frac{1}{4\pi }\left( \sum_{l=1,d-l={\rm odd}}^{d-1}+\sum_{l=d+1}^{M}%
\right) \sum_{m=0}^{l}U_{lm}B\left(
\frac{d}{2},\frac{l-d}{2}\right) B\left(
m+\frac{1}{2},\frac{d-1}{2}\right) . \label{ZasDI0}
\end{eqnarray}
As regards to $Z_{DI}^{(1)}(s)$, for $M\geq d$ the corresponding
expression (\ref{ZDI1as}) is finite at $s=-1$ and we can directly
put $\rho =0$. After integrating by parts and introducing polar
coordinates we find
\begin{equation}
Z_{DI}^{(1)}(-1)=\frac{1}{\pi }\int_{0}^{\infty
}dr\,r^{d-1}\int_{0}^{\pi /2}d\theta \,\sin ^{d-2}\theta \left[
\ln \left( \sqrt{2\pi r}
e^{-rg(\theta )}I_{r\cos \theta }(r\sin \theta )\right) -\sum_{l=1}^{M}\frac{%
U_{l}(\cos \theta )}{\left( 1+r^{2}\right) ^{l/2}}\right] .
\label{Z1DI-1}
\end{equation}
Collecting all terms for the expansion near $s=-1$ one finds
\begin{equation}
Z_{DI}(s)=\frac{Z_{DI,-1}^{(as)}}{s+1}%
+Z_{DI,0}^{(as)}+Z_{DI}^{(1)}(-1)+O(s+1),  \label{ZDIexpl}
\end{equation}
with the separate terms defined by (\ref{ZDIres}), (\ref
{ZasDI0}), (\ref{Z1DI-1}). Now the vacuum energy in the region
$0\leq \xi \leq \xi _1$ is the sum of the pole and finite terms
\begin{equation}\label{E1DL3}
    E_{1D}^{(L)}(\xi _1)=E_{1Dp}^{(L)}+E_{1Df}^{(L)},
\end{equation}
with
\begin{equation} \label{E1DLp}
E_{1Dp}^{(L)}= \frac{B_dZ_{DI,-1}^{(as)}}{\xi _1^{d-1} (s+1)},
\quad E_{1Df}^{(L)}= \frac{B_d}{\xi _1^{d-1} }
\left[Z_{DI,0}^{(as)}+Z_{DI}^{(1)}(-1)\right] .
\end{equation}
The results  of the numerical evaluations for these quantities are
given in Table \ref{tab:diren} for spatial dimensions $d=2,3,4$.
The corresponding vacuum energy $E_{1Dg}^{(L)}$ measured by an
uniformly accelerated observer is related to the energy
(\ref{E1DL3}) by formula (\ref{EgErel}). In particular, using the
data from Table \ref{tab:diren}, for the case of spatial dimension
$d=3$ one finds
\begin{equation}\label{EL1Dg}
E^{(L)}_{1Dg}(\xi _1)=\frac{g}{\xi _1^2} \left(
-0.000695-\frac{1}{1260\pi ^2}\left[ \frac{1}{s+1}+\ln \left(
\frac{\mu }{g}\right) \right] \right) ,
\end{equation}
where $g$ is the proper acceleration of the observer.

\subsection{Total Casimir energy for a single Dirichlet plate}
\label{subsec:totDir}

The total Casimir energy for a single Dirichlet plate can be
obtained by summing the RR and RL parts considered above:
\begin{equation}\label{totDiren1pl}
E_{1D}(\xi _1)=E_{1D}^{(R)}(\xi _1)+E_{1D}^{(L)}(\xi _1) .
\end{equation}
As separate summands, it contains pole and finite parts,
\begin{equation}\label{E1D3}
    E_{1D}(\xi _1)=E_{1Dp}+E_{1Df},
\end{equation}
where
\begin{equation} \label{E1Dp}
E_{1Dp}= \frac{B_d Z_{DK,-1}^{(as)}\left( 1+(-1)^d \right) }{\xi
_1^{d-1} (s+1)}, \quad E_{1Df}= \frac{B_d}{\xi _1^{d-1} }
\left[Z_{DK,0}^{(as)}+Z_{DI,0}^{(as)}+Z_{DK}^{(1)}(-1)+Z_{DI}^{(1)}(-1)\right]
.
\end{equation}
Note that in odd spatial dimensions the pole part vanishes due to
the cancellation of corresponding RR and RL parts and the total
Casimir energy is finite. In this case this energy can be
presented in the form
\begin{eqnarray}\label{E1Dtotal}
    E_{1D}&=& \frac{B_d}{\pi \xi _1^{d-1}}\left\{ \frac{1}{2}\sum_{l=1}^{M_1}
    B\left( \frac{d}{2},l-\frac{d}{2}\right) \sum_{m=0}^{2l}U_{2l,m}
    B\left( m+\frac{1}{2},\frac{d-1}{2}\right) +
    \int_{0}^{\infty }dr r^{d-1}\int_{0}^{\pi /2}d\theta \sin
    ^{d-2}\theta \right. \nonumber
    \\
    && \times \left. \left[ \ln \left( 2rI_{r\cos \theta }(r\sin \theta )
    K_{r\cos \theta }(r\sin \theta )\right) -2\sum_{l=1}^{M_1}
    \frac{U_{2l}(\cos \theta )}{(1+r^2)^l}\right] \right\} .
\end{eqnarray}
For the convergence of the integral in this formula it is
sufficient to take $M_1> d/2-1$. The numerical results for the
finite part of the total vacuum energy are given in Table
\ref{tab:diren}. For the total Casimir energy measured by an
uniformly accelerated observer with the proper acceleration $g$,
in odd dimensions the divergencies cancel and so do the
logarithmic terms. In particular, for $d=3$ Dirichlet scalar the
total Casimir energy
\begin{equation}\label{E1Dg}
E_{1Dg}=-\frac{0.00131 g}{\xi _{1}^{2}}
\end{equation}
is negative. This means that the resulting vacuum forces tend to
accelerate the Dirichlet plate.

\section{Casimir energy for two parallel Dirichlet
plates}\label{sec:Dir2pl}

\begin{figure}[tbph]
\begin{center}
\epsfig{figure=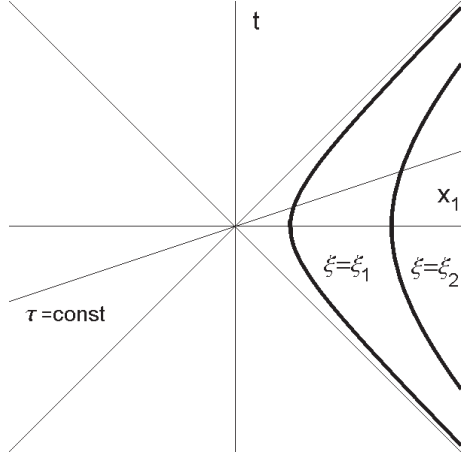,width=6cm,height=6cm}
\end{center}
\caption{The problem geometry in the case of two plates. The heavy
lines $\protect\xi =\protect\xi _{1}$ and $\protect\xi
=\protect\xi _{2}$ represent the trajectories of the plates.}
\label{fig22pl}
\end{figure}

In this section we consider the scalar vacuum in the region
between two plates located at $\xi =\xi _{1}$, $\xi =\xi _{2}$,
which therefore have proper accelerations $\xi _1^{-1}$ and $\xi
_2^{-1}$. The problem geometry is depicted in  Fig. \ref{fig22pl}.
The scalar field satisfies the Dirichlet boundary conditions on
the plates:
\begin{equation}
\varphi |_{\xi =\xi _{1}}=\varphi |_{\xi =\xi _{2}}=0.
\label{Dirbound2pl}
\end{equation}
The function $\phi (\xi )$ in Eq. (\ref{wavesracture}) satisfying
the boundary condition on the plate $\xi =\xi _2$ is in the form
\begin{equation}\label{Diom}
\phi (\xi )=D_{i\omega }(k\xi ,k\xi _{2})\equiv I_{i\omega }(k\xi
_{2})K_{i\omega }(k\xi )-I_{i\omega }(k\xi )K_{i\omega }(k\xi
_{2}).
\end{equation}
From the boundary conditions on the plate $\xi =\xi _1$ we find
that the corresponding eigenfrequencies are roots to the equation
\begin{equation}
D_{i\omega }(k\xi _{1},k\xi _{2})=0. \label{Dirmodes}
\end{equation}
This equation has an infinite set of solutions. We will denote the
positive roots by $\omega =\omega _{Dn}(k\xi _{1},k\xi _{2})$, and
will assume that they are arranged in the ascending order $\omega
_{Dn}<\omega _{Dn+1}$. For the vacuum energy in the region $\xi
_{1}\leq \xi \leq \xi _{2}$ per unit surface of the plates on has
\begin{equation}
E_{D}=\frac{1}{2}\int \frac{d^{d-1}k}{(2\pi
)^{d-1}}\sum_{n=1}^{\infty }\omega _{Dn}(k\xi _{1},k\xi _{2}).
\label{Diren}
\end{equation}
This energy can be written as
\begin{equation}\label{Diren1p}
    E_{D}=B_d
    \int _{0}^{\infty } dk \, k^{d-2} \zeta _D(s,k\xi _1,k\xi
    _2)|_{s=-1},
\end{equation}
with the partial zeta function
\begin{equation}
\zeta _{D}(s,k\xi _{1},k\xi _{2})=\sum_{n=1}^{\infty }\omega
_{Dn}^{-s}(k\xi _{1},k\xi _{2}),\quad {\mathrm{Re}}s>1.
\label{zetaD}
\end{equation}
In order to obtain the Casimir energy, one has to find the
analytic continuation of Eq. (\ref{zetaD}) to $s=-1$. This is done
by an analytic continuation of an adequate contour integration in
the complex plane. An immediate consequence of the Cauchy's
formula for the residues of a complex function is the expression
\begin{equation}
\zeta _{D}(s,k\xi _{1},k\xi _{2})=\frac{1}{2\pi i}\int_{C}dz\,z^{-s}\frac{%
\partial }{\partial z}\ln D_{iz}(k\xi _{1},k\xi _{2}),  \label{zetaD1}
\end{equation}
where the contour $C$ is the same as in Eq. (\ref{zet1D1}) with
the additional semicircle $C_{\rho }$ of small radius $\rho $
avoiding the origin from the right (in Eq. (\ref{zetaD1}) we can
directly put $\rho =0$, however this semicircle is needed for the
next step, see Eq. (\ref{zetaD2}) below). Let us denote by $C^{+}$
and $C^{-}$ the upper and lower halves of the contour $C $ except
the parts coming from the semicircle $C_{\rho }$. The integral on
the right of Eq. (\ref{zetaD1}) can be presented in the form
\begin{eqnarray}
\zeta _{D}(s,k\xi _{1},k\xi _{2}) &=&\zeta _{1D}^{(R)}(s,k\xi _{1})+\frac{1}{%
2\pi i}\sum_{\alpha =+,-}\int_{C^{\alpha }}dz\,z^{-s}\frac{\partial }{%
\partial z}\ln I_{-\alpha iz}(k\xi _{2})  \nonumber \\
&&+\frac{1}{2\pi i}\sum_{\alpha =+,-}\int_{C^{\alpha }}dz\,z^{-s}\frac{%
\partial }{\partial z}\ln \left[ 1-\frac{I_{-\alpha iz}(k\xi
_{1})K_{iz}(k\xi _{2})}{I_{-\alpha iz}(k\xi _{2})K_{iz}(k\xi
_{1})}\right]
\nonumber \\
&&+\frac{1}{2\pi i}\int_{C_{\rho }}dz\,z^{-s}\frac{\partial
}{\partial z}\ln D_{iz}(k\xi _{1},k\xi _{2}), \label{zetaD2}
\end{eqnarray}
where we have used (\ref{zet1D1}) to introduce the function $\zeta
_{1D}^{(R)}(s,x)$. The last integral over $C_{\rho }$ vanishes in
the limit $\rho \rightarrow 0$ for ${\mathrm{Re}}s<2$ and we will
omit it in the following consideration. After parametrizing the
integrals over the imaginary axis, we arrive at the formula
\begin{eqnarray}
\zeta _{D}(s,k\xi _{1},k\xi _{2}) &=&\zeta _{1D}^{(R)}(s,k\xi _{1})+\frac{1}{%
\pi }\sin \frac{\pi s}{2}\int_{\rho }^{\infty }dz\,z^{-s}\frac{\partial }{%
\partial z}\ln I_{z}(k\xi _{2})  \label{zetaD3} \\
&&+\frac{1}{\pi }\sin \frac{\pi s}{2}\int_{\rho }^{\infty }dz\,z^{-s}\frac{%
\partial }{\partial z}\ln \left[ 1-\frac{I_{z}(k\xi _{1})K_{z}(k\xi _{2})}{%
I_{z}(k\xi _{2})K_{z}(k\xi _{1})}\right] .  \nonumber
\end{eqnarray}
The last integral on the right of this formula is finite at $s=-1$
and vanishes in the limits $\xi _{1}\rightarrow 0$ or $\xi
_{2}\rightarrow \infty $. It follows from here that the second
term on the right corresponds to the zeta function $\zeta
_{1D}^{(L)}(s,k\xi _{2})$ for the region on the left of a single
plate located at $\xi =\xi _{2}$. The analytic continuation of
this function we have done in the previous section.

\begin{figure}[tbph]
\begin{center}
\epsfig{figure=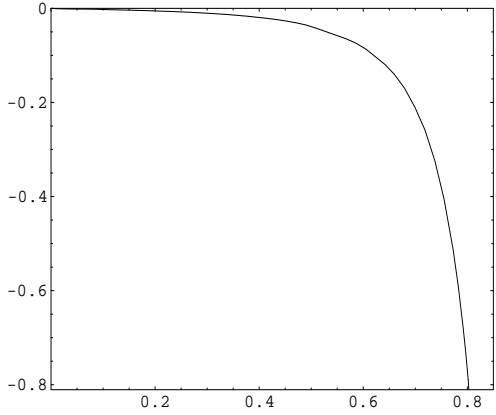,width=6.5cm,height=5.5cm}
\end{center}
\caption{Interference part of the Casimir energy in the region
between two Dirichlet plates, $\xi _2^{d-1}\Delta E_D(\xi _1,\xi
_2)$, as a function on the ratio $\xi _1/\xi _2$ for $d=3$.}
\label{fig2intd}
\end{figure}

As a result the vacuum energy in the region $\xi _1\leq \xi \leq
\xi _2$ can be written  in the form of the sum
\begin{equation}\label{ED2pl}
    E_D=E_{1D}^{(R)}(\xi _1)+E_{1D}^{(L)}(\xi _2)+\Delta E_D(\xi _1,\xi
    _2),
\end{equation}
where the interference term is given by the formula
\begin{equation}\label{EDint}
\Delta E_D(\xi _1,\xi _2)=\frac{B_d}{\pi } \int_{0}^{\infty }dk \,
k^{d-2} \int_{0}^{\infty }dz\, \ln \left[ 1-\frac{I_z(k \xi
_1)K_z(k \xi _2)}{I_z(k \xi _2)K_z(k \xi _1)}\right] .
\end{equation}
Note that in the corresponding expression we have partially
integrated the $z$-integral. The function $D_z(k\xi _1,k\xi _2)$
is positive for $\xi _1<\xi _2$ and the energy (\ref{EDint}) is
always negative, $\Delta E_D(\xi _1,\xi _2)<0$. In Fig.
\ref{fig2intd} we have presented the dependence of the
interference part of the Casimir energy on the ratio $\xi _1/\xi
_2$ for $d=3$.
In the case of two plates geometry, to obtain the total Casimir
energy, $E_{D}^{{\mathrm{(tot)}}}$, we need to add to the energy
in the region between the plates, given by Eq. (\ref{ED2pl}), the
energies coming from the regions $\xi \leq \xi _1$ and $\xi \geq
\xi _2$. As a result one receives
\begin{equation}\label{ED2pltot}
E_D^{{\mathrm{(tot)}}}=E_{1D}(\xi _1)+E_{1D}(\xi _2)+\Delta
E_D(\xi _1,\xi _2),
\end{equation}
where the interference part is given by formula (\ref{EDint}). Note that in the limit $\xi _2 \to \infty $ and for a fixed $\xi _1$, the last two terms on the right vanish and we recover the result for a single plate located at $\xi =\xi _1$. For $d=3 $ Dirichlet scalar all terms on the right of formula (\ref{ED2pltot}) are negative and, hence, the total Casimir energy for two parallel plates is negative.

In the limit $\xi _{1}\to \xi _{2}$, expression (\ref{EDint}) is
divergent and for small values of $\xi _{2}/\xi _{1}-1$ the main
contribution comes from large values of $z$. Replacing the order
of integrations and introducing a new integration variable
$x=k/z$, we can replace the Bessel modified functions by their
uniform asymptotic expansions for large values of the order. After
evaluating the integrals, to the leading order one receives
\begin{equation}\label{DelEDlim1}
\Delta E_D(\xi _1,\xi _2) \sim -(4\pi )^{-\frac{d+1}{2}}\zeta
_R(d+1)\Gamma \left( \frac{d+1}{2}\right) \frac{\xi _1}{(\xi
_2-\xi _1)^d},
\end{equation}
where $\zeta _{R}(x)$ is the Riemann zeta function. Note that the
energy $\Delta E_D$ is related to the corresponding quantity
$\Delta E_{Dg}$ measured by an uniformly accelerated observer by
formula $\Delta E_{Dg}=g \Delta E_D$, where $g$ is the proper
acceleration of the observer (see Appendix \ref{ap:propE}). In the
limit $\xi _1, \xi _2\to \infty $, $\xi _2-\xi
_1={\mathrm{const}}$, the parts in Eq. (\ref{ED2pltot})
corresponding to the contributions from the single plates vanish,
$E_{1D}(\xi _i)\to 0$, $i=1,2$, and the interference part remains
only. In this case, as it follows from Eq. (\ref{DelEDlim1}), the
energy measured by an observer with the acceleration $g\to 0$ such
that $g\xi _1\to 1$, coincides with the standard Casimir energy
for two parallel plates in $d+1$-dimensional Minkowski spacetime
\cite{Ambj83}.

As it has been shown in Ref. \cite{Avag02}, the vacuum forces
acting on the boundaries can be presented as a sum of the
self-action and interaction forces.  It can be easily checked that
the interaction forces are related to the energy (\ref{EDint}) by
the standard thermodynamical relations
\begin{subequations}\label{pD1int}
\begin{eqnarray}
\frac{\partial }{\partial \ln \xi _1}\Delta E_D(\xi _1,\xi _2)&=&
p_{D{\mathrm{(int)}}}^{(1)}(\xi _1,\xi _2)=-\frac{B_d}{\pi \xi
_1^2} \int_{0}^{\infty }dk \, k^{d-2} \int_{0}^{\infty }dz\,
\frac{K_z(k
\xi _2)}{K_z(k \xi _1)D_z(k\xi _1, k\xi _2)} \\
\frac{\partial }{\partial \ln \xi _2}\Delta E_D(\xi _1,\xi _2)&=&
-p_{D{\mathrm{(int)}}}^{(2)}(\xi _1,\xi _2)=\frac{B_d}{\pi \xi
_2^2} \int_{0}^{\infty }dk \, k^{d-2} \int_{0}^{\infty }dz\,
\frac{I_z(k \xi _1)}{I_z(k \xi _2)D_z(k\xi _1, k\xi _2)} ,
\end{eqnarray}
\end{subequations}
where $p_{D{\mathrm{(int)}}}^{(j)}$, $j=1,2$ are the vacuum
effective pressures on the plate at $\xi =\xi _j$ induced by the
presence of the second plate (interaction force per unit surface).

\section{Casimir energy for a single plate with Neumann boundary
condition}\label{sec:neuscal}

\subsection{RR region}\label{subsec:neuRR}

In this section we will consider the vacuum energy for a scalar
field satisfying Neumann boundary condition on a single plate
located at $\xi =\xi _{1}$:
\begin{equation}
\frac{\partial \varphi }{\partial \xi }|_{\xi =\xi _{1}}=0.
\label{NR1pl}
\end{equation}
In the region $\xi \geq \xi _{1}$, the corresponding
eigenfrequencies are roots to the equation
\begin{equation}
K_{i\omega }^{\prime }(k\xi _{1})=0,  \label{Kderiom}
\end{equation}
where the prime denotes the derivative with respect to the
argument of the function. The positive roots to this equation
arranged in the ascending order we will denote by $\omega =\omega
_{1Nn}(k\xi _{1})$, $n=1,2,\ldots $, $\omega _{1Nn}<\omega
_{1Nn+1}$. The vacuum energy per unit surface of the plate in the
region $\xi \geq \xi _{1}$ is given by formula
\begin{equation}
E_{1N}^{(R)}(\xi _{1})=\frac{B_d}{\xi _1^{d-1}}\int_{0}^{\infty
}dx\,x^{d-2}\sum_{n=1}^{\infty }\omega _{1Nn}(x).  \label{E1NR}
\end{equation}
To regularize this energy we consider the function
\begin{equation}
Z_{NK}(s)=\int_{0}^{\infty }dx\,x^{d-2}\zeta _{1N}^{(R)}(s,x),
\label{Zs0N}
\end{equation}
with the partial zeta function related to the corresponding
eigenfrequences:
\begin{equation}
\zeta _{1N}^{(R)}(s,x)=\sum_{n=1}^{\infty }\omega _{1Nn}^{-s}(x).
\label{zet1N}
\end{equation}
The Casimir energy is obtained by the analytic continuation to
$s=-1$:
\begin{equation}
E_{1N}^{(R)}(\xi _{1})=\frac{B_d}{\xi _1^{d-1} }Z_{NK}(s)|_{s=-1}.
\label{EnZsN}
\end{equation}
This analytic continuation procedure is similar to that presented
above for the Dirichlet case. First, we present the partial zeta
function in terms of a contour integral
\begin{equation}
\zeta _{1N}^{(R)}(s,x)=\frac{1}{2\pi i}\int_{C}dz\,z^{-s}\frac{\partial }{%
\partial z}\ln K_{iz}^{\prime }(x),  \label{zet1N1}
\end{equation}
with the same contour $C$ as in Eq. (\ref{zet1D}). When the radius
of the large semicircle of this contour tends to infinity the
corresponding contribution into $\zeta _{1N}^{(R)}(s,x)$ vanishes
for ${\mathrm{Re}}s>1$. After parametrizing the integrals over
imaginary axis we obtain the following integral representation
\begin{equation}
\zeta _{1N}^{(R)}(s,x)=\frac{1}{\pi }\sin \frac{\pi s}{2}\int_{0
}^{\infty }dz\,z^{-s}\frac{\partial }{\partial z}\ln K_{z}^{\prime
}(x),  \label{zet1N2}
\end{equation}
valid for $1<{\mathrm{Re}}s<2$. For the analytic continuation of
Eq. (\ref{zet1N2}) to $s=-1$ we use the uniform asymptotic
expansion of the function $K'_{z}(x)$ for large values of the
order. We will write this expansion in the form
\begin{equation}
K_{z}^{\prime }(x)=-\sqrt{\frac{\pi }{2}}\frac{\left(
x^{2}+z^{2}\right) ^{1/4}}{x}e^{-z\eta (x/z)}K_{z}^{(N)}(x),\quad
K_{z}^{(N)}(x)\sim \sum_{l=0}^{\infty
}(-1)^{l}\frac{\overline{v}_{l}(t)}{\left( x^{2}+z^{2}\right)
^{l/2}},  \label{Kderzas}
\end{equation}
where
\begin{equation}
\overline{v}_{l}(t)=\frac{v_{l}(t)}{t^{l}},  \label{tuN}
\end{equation}
and the expressions of the functions $v_{l}(t)$ can be found in
Ref. \cite{Abramowitz}. From these expressions we can see that the
coefficients $ \overline{v}_{l}(t)$ are polynomials in $t$:
\begin{equation}
\overline{v}_{l}(t)=\sum_{m=0}^{l}v_{lm}t^{2m}.  \label{vlbar}
\end{equation}
From the formulae for the functions $v_{l}(t)$ (see
\cite{Abramowitz}), the coefficients $v_{lm}$ can be expressed via
the coefficients $u_{lm}$ in Eq. (\ref{ulbar}) by the relations
\begin{equation}
v_{l+1,m}=-\frac{1}{2}u_{lm}\left[ 2m+l+1-\frac{1}{4(2m+l+1)}\right] +\frac{1%
}{2}u_{l,m-1}\left[ 2m+l-1-\frac{5}{4(2m+l+1)}\right] ,
\label{recvl}
\end{equation}
where $m=0,1,\ldots ,l+1$, and, as has been noted before, $%
u_{l,-1}=u_{l,l+1}=0$, $u_{00}=1$. Using expansion
(\ref{Kderzas}), the zeta function may be split into two pieces
\begin{equation}
\zeta _{1N}^{(R)}(s,x)=\zeta _{1N}^{(R0)}(s,x)+\zeta
_{1N}^{(R1)}(s,x), \label{zet1N3}
\end{equation}
with
\begin{eqnarray}
\zeta _{1N}^{(R0)}(s,x) &=&\frac{1}{\pi }\sin \frac{\pi
s}{2}\int_{0
}^{\infty }dz\,z^{-s}\frac{\partial }{\partial z}\ln \sqrt{\frac{\pi }{2}}%
\frac{\left( x^{2}+z^{2}\right) ^{1/4}}{x}e^{-z\eta (x/z)},
\label{zet1N01}
\\
\zeta _{1N}^{(R1)}(s,x) &=&\frac{1}{\pi }\sin \frac{\pi
s}{2}\int_{0}^{\infty }dz\,z^{-s}\frac{\partial }{\partial z}\ln
K_{z}^{(N)}(x). \label{zet1N11}
\end{eqnarray}
Under the conditions $1<{\mathrm{Re}}s<2$, evaluating the integral
in Eq. (\ref{zet1N01}) one finds
\begin{equation}
\zeta _{1N}^{(R0)}(s,x)=\frac{(x/2)^{1-s}}{\pi (1-s)}B\left( 1-s,\frac{s-1}{2}%
\right) \sin \frac{\pi s}{2}+\frac{x^{-s}}{4}.  \label{zet1N02}
\end{equation}
By the arguments similar to those for the Dirichlet case, we
conclude that the contribution of the term $\zeta
_{1N}^{(R0)}(s,x)$ into Eq. (\ref{Zs0N}) vanishes. The expression
with the second term on the right of Eq. (\ref{zet1N3}) takes the
form
\begin{equation}
Z_{NK}(s)=\frac{1}{\pi }\sin \frac{\pi s}{2}\int_{0}^{\infty
}dx\,x^{d-2}\int_{0}^{\infty }dz\,z^{-s}\frac{\partial }{\partial
z}\ln K_{z}^{(N)}(x).  \label{ZsN}
\end{equation}
Now we subtract and add to the integrand in Eq. (\ref{ZsN}) the
corresponding asymptotic expression and exactly integrate the
asymptotic part. For this we introduce polar coordinates in the
$(z,x)$ plane and the coefficients $V_{l}(\cos \theta )$ in
accordance with
\begin{equation}
\ln \left( \sum_{l=0}^{\infty }(-1)^{l}\frac{\overline{v}_{l}(\cos \theta )}{%
r^{l}}\right) =\sum_{l=1}^{\infty }(-1)^{l}\frac{V_{l}(\cos \theta
)}{\left( 1+r^{2}\right) ^{l/2}}.  \label{defVl}
\end{equation}
This allows us to present (\ref{ZsN}) in the form of the sum
\begin{equation}
Z_{NK}(s)=Z_{NK}^{(as)}(s)+Z_{NK}^{(1)}(s),  \label{Zs1N}
\end{equation}
where
\begin{eqnarray}
Z_{NK}^{(as)}(s) &=&\frac{1}{\pi }\sin \frac{\pi
s}{2}\int_{0}^{\infty
}dx\,x^{d-2}\int_{0}^{\infty }dz\,z^{-s}\frac{\partial }{\partial z}%
\sum_{l=1}^{M}(-1)^{l}\frac{V_{l}(\cos \theta )}{\left( 1+r^{2}\right) ^{l/2}%
},  \label{Zsas1N} \\
Z_{NK}^{(1)N}(s) &=&\frac{1}{\pi }\sin \frac{\pi
s}{2}\int_{0}^{\infty
}dx\,x^{d-2}\int_{0}^{\infty }dz\,z^{-s}\frac{\partial }{\partial z}%
\left[ \ln K_{z}^{(N)}(x)-\sum_{l=1}^{M}(-1)^{l}\frac{V_{l}(\cos \theta )}{%
\left( 1+r^{2}\right) ^{l/2}}\right] .  \label{Zs11N}
\end{eqnarray}
For $M\geq d$ the expression $Z_{NK}^{(1)}(s)$ is finite at $s=-1$
and we can perform the integral numerically. For numerical
evaluation it is useful to integrate by parts the $z$-integral and
to perform to the polar coordinates. This yields to the final
expression
\begin{eqnarray}
Z_{NK}^{(1)}(-1) &=& \frac{1}{\pi }\int_{0}^{\infty
}dr\,r^{d-1}\int_{0}^{\pi /2}d\theta \,\sin ^{d-2}\theta \nonumber
\\
&& \times \left[ \ln \left( -\sqrt{\frac{2r}{\pi }}\sin \theta
e^{rg(\theta )}K_{r\cos \theta }^{\prime }(r\sin \theta )\right)
-\sum_{l=1}^{M}(-1)^{l}\frac{V_{l}(\cos \theta )}{\left(
1+r^{2}\right) ^{l/2}}\right] .  \label{ZNK1-1}
\end{eqnarray}

Now let us consider the asymptotic part of the function
$Z_{NK}(s)$ given by expression (\ref{Zsas1N}). To evaluate this
part note that the functions $V_{l}(t)$ have the structure
\begin{equation}
V_{l}(t)=\sum_{m=0}^{l}V_{lm}t^{2m},  \label{Vlm}
\end{equation}
where the numerical coefficients $V_{lm}$ are related to the
coefficients in Eq. (\ref{vlbar}) by expansion (\ref{defVl}). The
first five functions are
\begin{subequations}\label{V123}
\begin{eqnarray}
V_{1}(t) &=&-\frac{3}{8}+\frac{7}{24}t^{2},\quad V_{2}(t)=-\frac{3}{16}+%
\frac{5}{8}t^{2}-\frac{7}{16}t^{4},   \\
V_{3}(t) &=&-\frac{45}{128}+\frac{2887}{1920}t^{2}-\frac{315}{128}t^{4}+%
\frac{1463}{1152}t^{6},   \\
V_{4}(t) &=&-\frac{51}{128}+\frac{129}{32}t^{2}-\frac{761}{64}t^{4}+\frac{441%
}{32}t^{6}-\frac{707}{128}t^{8},   \\
V_{5}(t) &=&-\frac{3879}{5120}+\frac{436931}{35840}t^{2}-\frac{264745}{4608}%
t^{4}+\frac{173209}{1536}t^{6}-\frac{101395}{1024}t^{8}+\frac{495271}{15360}%
t^{10}.
\end{eqnarray}
\end{subequations}
Integrating over $z$ by parts and introducing polar coordinates,
for the asymptotic part one finds
\begin{equation}
Z_{NK}^{(as)}(s)=\frac{s}{\pi }\sin \frac{\pi s}{2}\sum_{l=1}^{M}(-1)^{l}%
\sum_{m=0}^{l}V_{lm}\int_{0}^{\infty }dr\frac{r^{d-s-2}}{\left(
1+r^{2}\right) ^{l/2}}\int_{0}^{\pi /2}d\theta \sin ^{d-2}\theta
\cos ^{2m-s-1}\theta .  \label{Zsas2N}
\end{equation}
The evaluation of the integrals leads to the following formula
\begin{equation}
Z_{NK}^{(as)}(s)=\frac{s}{4\pi }\sin \frac{\pi s}{2}\sum_{l=1}^{M}(-1)^{l}%
\sum_{m=0}^{l}V_{lm}B\left(
\frac{d-s-1}{2},\frac{l+s-d+1}{2}\right) B\left(
m-\frac{s}{2},\frac{d-1}{2}\right) .  \label{Zsas3N}
\end{equation}
Due to the first beta function on the right this expression has a
simple pole at $s=-1$ with the residue
\begin{equation}
Z_{NK,-1}^{(as)}=\frac{(-1)^{d}}{\pi d}\sum_{p=0}^{p_{d}}(-1)^{p}%
\sum_{m=0}^{d-2p}V_{d-2p,m}\frac{B\left( m+1/2,(d-1)/2\right) }{B(d/2-p,p+1)}%
,  \label{ZsresN}
\end{equation}
where $p_d$ is defined in Eq. (\ref{pd}). Hence, the Laurent
expansion of the asymptotic part near $s=-1$ has the form
\begin{equation}
Z_{NK}^{(as)}(s)=\frac{Z_{NK,-1}^{(as)}}{s+1}+Z_{NK,0}^{(as)}+O(s+1),
\label{Zsas4N}
\end{equation}
with the finite part
\begin{eqnarray}
Z_{NK,0}^{(as)} &=&\frac{(-1)^{d}}{2\pi d}\sum_{p=0}^{p_{d}}(-1)^{p}%
\sum_{m=0}^{d-2p}V_{d-2p,m}\frac{B\left(
m+\frac{1}{2},\frac{d-1}{2}\right) }{B(\frac{d}{2}-p,p+1)}
\nonumber \\
&& \times \left[ \psi (p+1)-\psi \left( \frac{d}{2}\right) +\psi
\left( m+\frac{d}{2}\right) -\psi \left( m+\frac{1}{2}\right)
-2\right]
\nonumber \\
&&+\frac{1}{4\pi }\left( \sum_{l=1,d-l={\rm odd}}^{d-1}+\sum_{l=d+1}^{M}%
\right) (-1)^{l}\sum_{m=0}^{l}V_{lm}B\left(
\frac{d}{2},\frac{l-d}{2}\right) B\left(
m+\frac{1}{2},\frac{d-1}{2}\right) , \label{ZasNI0}
\end{eqnarray}
where the second sum in the braces of the third line is present only for $%
M\geq d+1$. Now adding the part coming from (\ref{ZNK1-1}), for
the Laurent expansion of the function $Z_{NK}(s)$ near $s=-1$ one
finds
\begin{equation}
Z_{NK}(s)=\frac{Z_{NK,-1}^{(as)}}{s+1}%
+Z_{NK,0}^{(as)}+Z_{NK}^{(1)}(-1)+O(s+1),  \label{ZNKexp}
\end{equation}
where the separate terms are defined by formulae (\ref{ZNK1-1}),
(\ref {ZsresN}), (\ref{ZasNI0}). Using this result, for the vacuum
energy in the case of Neumann scalar induced by a single plate at
$\xi =\xi _1$ in the region $\xi \geq \xi _1$ one receives
\begin{equation}\label{E1NR3}
    E_{1N}^{(R)}(\xi _1)=E_{1Np}^{(R)}+E_{1Nf}^{(R)},
\end{equation}
where for the pole and finite contributions we have
\begin{equation}\label{E1NRp}
E_{1Np}^{(R)}= \frac{B_dZ_{NK,-1}^{(as)}}{\xi _1^{d-1} (s+1)},
\quad E_{1Nf}^{(R)}= \frac{B_d}{\xi _1^{d-1} }
\left[Z_{NK,0}^{(as)}+Z_{NK}^{(1)}(-1)\right] .
\end{equation}
The numerical results corresponding to the vacuum energy
(\ref{E1NR3}) with separate pole and finite parts are presented in
Table \ref{tab:neumen} for spatial dimensions $d=2,3,4$. Comparing
to the data from Table \ref{tab:diren}, we see that in all these
cases the Neumann quantities for the RR region dominate the
Dirichlet ones.

\begin{table}
  \centering
  \caption{Pole and finite parts of the Casimir energy for a single
  Neumann plate.}\label{tab:neumen}
  \begin{tabular}{cccccc}
\hline \hline
  $d$ & $\xi _1^{d-1}E_{1Np}^{(R)}$ & $\xi _1^{d-1}E_{1Nf}^{(R)}$ &
  $\xi _1^{d-1}E_{1Np}^{(L)}$ & $\xi _1^{d-1}E_{1Nf}^{(L)}$&
  $\xi _1^{d-1}E_{1Nf}$ \\
  \hline
  2 & $-\frac{5}{512\pi (s+1)}$ & -0.00874 & $-\frac{5}{512\pi (s+1)}$ & -0.000633 & -0.00937 \\

  3 & $\frac{1}{180\pi ^2(s+1)}$ & 0.00213 & $-\frac{1}{180\pi ^2(s+1)} $& 0.000792 &
  0.00292  \\

  4 & $-\frac{269}{262144\pi ^2 (s+1)}$ & -0.000593 & $-\frac{269}{262144\pi ^2 (s+1)}$ &
  -0.000398 & -0.000194  \\
  \hline
  \hline
\end{tabular}
\end{table}
By making use the formula given in Appendix \ref{ap:propE}, we can
obtain the vacuum energy in the RL region measured by an uniformly
accelerated observer. In $d=3$ spatial dimension for this energy
one has
\begin{equation}\label{ER1Ng}
E^{(R)}_{1Ng}(\xi _1)=\frac{g}{\xi _1^2} \left(
0.00213+\frac{1}{180\pi ^2}\left[ \frac{1}{s+1}+\ln \left(
\frac{\mu }{g}\right) \right] \right) .
\end{equation}
It logarithmically depends on the normalization scale.

\subsection{Neumann vacuum energy in the RL region}\label{subsec:neuRL}

Similar to the case of the RR region, the Casimir energy for the
Neumann scalar in the region between a single plate at $\xi =\xi
_{1}$ and the Rindler horizon corresponding to $\xi =0$ is
determined by the formula (see Sec. \ref{sec:Neu2pl} below)
\begin{equation}
E_{1N}^{(L)}(\xi _{1})=\frac{B_d}{\xi _1^{d-1} }Z_{NI}(s)|_{s=-1},
\label{EL1N}
\end{equation}
with the function
\begin{equation}
Z_{NI}(s)=\int_{0}^{\infty }dx\,x^{d-2}\zeta _{1N}^{(L)}(s,x),
\label{ZNlI}
\end{equation}
and the partial zeta function
\begin{equation}
\zeta _{1N}^{(L)}(s,x)=\frac{1}{\pi }\sin \frac{\pi
s}{2}\int_{\rho }^{\infty }dz\,z^{-s}\frac{\partial }{\partial
z}\ln I_{z}^{\prime }(x). \label{zetaLN}
\end{equation}
For the analytic continuation of the function $Z_{NI}(s)$ to
$s=-1$ we note that for the derivative of the Bessel modified
function one has the uniform asymptotic expansion
\begin{equation}
I_{z}^{\prime }(x)=\frac{1}{\sqrt{2\pi }}\frac{\left(
x^{2}+z^{2}\right) ^{1/4}}{x}e^{z\eta (x/z)}I_{z}^{(N)}(x),\quad
I_{z}^{(N)}(x)\sim \sum_{l=0}^{\infty
}\frac{\overline{v}_{l}(t)}{\left( x^{2}+z^{2}\right) ^{l/2}},
\label{IzasN}
\end{equation}
where the functions $\eta (x)$, $t$ are defined in Eq. (\ref{tu}).
Separating the contributions coming from the different factors in
Eq. (\ref{IzasN}), we present the partial zeta function in the
form
\begin{equation}
\zeta _{1N}^{(L)}(s,x)=\zeta _{1N}^{(L0)}(s,x)+\zeta
_{1N}^{(L1)}(s,x), \label{zet1LN}
\end{equation}
with
\begin{eqnarray}
\zeta _{1N}^{(L0)}(s,x) &=&\frac{1}{\pi }\sin \frac{\pi
s}{2}\int_{\rho
}^{\infty }dz\,z^{-s}\frac{\partial }{\partial z}\ln \frac{1}{\sqrt{2\pi }}%
\frac{\left( x^{2}+z^{2}\right) ^{1/4}}{x}e^{z\eta (x/z)},
\label{zetL1N01}
\\
\zeta _{1N}^{(L1)}(s,x) &=&\frac{1}{\pi }\sin \frac{\pi
s}{2}\int_{\rho }^{\infty }dz\,z^{-s}\frac{\partial }{\partial
z}\ln I_{z}^{(N)}(x). \label{zetL1N11}
\end{eqnarray}
For $1<{\mathrm{Re}}s<2$ the $z$-integral in the expression for
the function $\zeta _{1N}^{(L0)}(s,x)$ is convergent in the limit
$\rho \rightarrow 0$ and the straightforward evaluation of the
integral in this limit gives
\begin{equation}
\zeta _{1N}^{(L0)}(s,x)=-\frac{(x/2)^{1-s}}{\pi (1-s)}B\left( 1-s,\frac{s-1}{2}%
\right) \sin \frac{\pi s}{2}+\frac{x^{-s}}{4}.  \label{zetL1N02}
\end{equation}
As in the case of RR region, the contribution of this term into
the renormalized vacuum energy vanishes. Hence, we need to
consider the quantity
\begin{equation}
Z_{NI}(s)=\frac{1}{\pi }\sin \frac{\pi s}{2}\int_{0}^{\infty
}dx\,x^{d-2}\int_{0}^{\infty }dz\,z^{-s}\frac{\partial }{\partial
z}\ln I_{z}^{(N)}(x).  \label{ZNI}
\end{equation}
With the help of asymptotic expansion (\ref{IzasN}), it can be
presented in the form
\begin{equation}
Z_{NI}(s)=Z_{NI}^{(as)}(s)+Z_{NI}^{(1)}(s),  \label{ZNI1}
\end{equation}
where
\begin{eqnarray}
Z_{NI}^{(as)}(s) &=&\frac{1}{\pi }\sin \frac{\pi
s}{2}\int_{0}^{\infty
}dx\,x^{d-2}\int_{\rho }^{\infty }dz\,z^{-s}\frac{\partial }{\partial z}%
\sum_{l=1}^{M}\frac{V_{l}(\cos \theta )}{\left( 1+r^{2}\right)
^{l/2}},
\label{ZNI1as} \\
Z_{NI}^{(1)}(s) &=&\frac{1}{\pi }\sin \frac{\pi
s}{2}\int_{0}^{\infty
}dx\,x^{d-2}\int_{\rho }^{\infty }dz\,z^{-s}\frac{\partial }{\partial z}%
\left( \ln I_{z}^{(N)}(x)-\sum_{l=1}^{M}\frac{V_{l}(\cos \theta
)}{\left( 1+r^{2}\right) ^{l/2}}\right) .  \label{ZNI11}
\end{eqnarray}
Note that for $M\geq d$ the expression (\ref{ZNI11}) for
$Z_{NI}^{(1)}(s)$ is finite at $s=-1$ and we can directly put
$\rho =0$. After integrating by parts and introducing polar
coordinates we find
\begin{equation}
Z_{NI}^{(1)N}(-1)=\frac{1}{\pi }\int_{0}^{\infty
}dr\,r^{d-1}\int_{0}^{\pi
/2}d\theta \,\sin ^{d-2}\theta \left[ \ln \left( \sqrt{2\pi r}%
e^{-rg(\theta )}\sin \theta I_{r\cos \theta }^{\prime }(r\sin
\theta )\right) -\sum_{l=1}^{M}\frac{V_{l}(\cos \theta )}{\left(
1+r^{2}\right) ^{l/2}}\right] .  \label{Z1NI-1}
\end{equation}
As regards the asymptotic part, in the expression for
$Z_{NI}^{(as)}(s)$ we can directly put $\rho =0$ and after
introducing polar coordinates and integrating one finds
\begin{equation}
Z_{NI}^{(as)}(s)=\frac{s}{4\pi }\sin \frac{\pi s}{2}\sum_{l=1}^{M}%
\sum_{m=0}^{l}V_{lm}B\left(
\frac{d-s-1}{2},\frac{l+s-d+1}{2}\right) B\left(
m-\frac{s}{2},\frac{d-1}{2}\right) .  \label{ZNIas2}
\end{equation}
At $s=-1$ this function has a simple pole with the residue
\begin{equation}
Z_{NI,-1}^{(as)}= (-1)^dZ_{NK,-1}^{(as)},  \label{ZNIres}
\end{equation}
with $Z_{NK,-1}^{(as)}$ defined by Eq. (\ref{ZsresN}). Now, taking
into account the contribution coming from (\ref{ZNI11}), we have
the following Laurent expansion near $s=-1$
\begin{equation}
Z_{NI}(s)=\frac{Z_{NI,-1}^{(as)}}{s+1}%
+Z_{NI,0}^{(as)}+Z_{NI}^{(1)}(-1)+O(s+1),  \label{ZNIexpl}
\end{equation}
where
\begin{eqnarray}
Z_{NI,0}^{(as)} &=&\frac{1}{2\pi d}\sum_{p=0}^{p_{d}}(-1)^{p}%
\sum_{m=0}^{d-2p}V_{d-2p,m}\frac{B\left(
m+\frac{1}{2},\frac{d-1}{2}\right) }{B(\frac{d}{2}-p,p+1)}
\nonumber \\
&& \times \left[ \psi (p+1)-\psi \left( \frac{d}{2}\right) +\psi
\left( m+\frac{d}{2}\right) -\psi \left( m+\frac{1}{2}\right)
-2\right]
\nonumber \\
&&+\frac{1}{4\pi }\left( \sum_{l=1,d-l={\rm odd}}^{d-1}+\sum_{l=d+1}^{M}%
\right) \sum_{m=0}^{l}V_{lm}B\left(
\frac{d}{2},\frac{l-d}{2}\right) B\left(
m+\frac{1}{2},\frac{d-1}{2}\right) \label{ZasNI01}.
\end{eqnarray}

The vacuum energy for the Neumann scalar in the region $0\leq \xi
\leq \xi _1$ is presented as a sum of the pole and finite terms
\begin{equation}\label{E1NL3}
    E_{1N}^{(L)}(\xi _1)=E_{1Np}^{(L)}+E_{1Nf}^{(L)},
\end{equation}
with
\begin{equation} \label{E1NLp}
E_{1Np}^{(L)}= \frac{B_dZ_{NI,-1}^{(as)}}{\xi _1^{d-1} (s+1)},
\quad E_{1Nf}^{(L)}= \frac{B_d}{\xi _1^{d-1} }
\left[Z_{NI,0}^{(as)}+Z_{NI}^{(1)}(-1)\right] .
\end{equation}
For spatial dimensions $d=2,3,4$ the results  of the corresponding
numerical evaluations are reported in Table \ref{tab:neumen}. For
spatial dimension $d=3$, the Casimir energy in the RL region
measured by an uniformly accelerated observer is obtained with the
help of formula (\ref{EgErel}):
\begin{equation}\label{EL1Ng}
E^{(L)}_{1Ng}(\xi _1)=\frac{g}{\xi _1^2} \left(
-0.000792-\frac{1}{180\pi ^2}\left[ \frac{1}{s+1}+\ln \left(
\frac{\mu }{g}\right) \right] \right) ,
\end{equation}
where $g$ is the proper acceleration for the observer.

\subsection{Total Casimir energy for a single Neumann plate}
\label{subsec:totNeu}

Summing the contributions from the RR and RL regions we obtain the
total Casimir energy for a single Neumann plate:
\begin{equation}\label{totNeuen1pl}
E_{1N}(\xi _1)=E_{1N}^{(R)}(\xi _1)+E_{1N}^{(L)}(\xi _1) .
\end{equation}
It can be divided into pole and finite parts,
\begin{equation}\label{E1N3}
    E_{1N}(\xi _1)=E_{1Np}+E_{1Nf},
\end{equation}
where
\begin{equation} \label{E1Np}
E_{1Np}= \frac{B_dZ_{NK,-1}^{(as)}\left( 1+(-1)^d\right) }{\xi
_1^{d-1} (s+1)}, \quad E_{1Nf}= \frac{B_d}{\xi _1^{d-1} }
\left[Z_{NK,0}^{(as)}+Z_{NI,0}^{(as)}+Z_{NK}^{(1)}(-1)+Z_{NI}^{(1)}(-1)\right]
.
\end{equation}
In odd spatial dimensions the pole part vanishes due to the
cancellation of corresponding RR and RL parts and the total
Casimir energy is finite. In this case this energy can be
presented in the form
\begin{eqnarray}\label{E1DNtotal}
    E_{1N}&=& \frac{B_d}{\pi \xi _1^{d-1}}\left\{ \frac{1}{2}\sum_{l=1}^{M_1}
    B\left( \frac{d}{2},l-\frac{d}{2}\right) \sum_{m=0}^{2l}V_{2l,m}
    B\left( m+\frac{1}{2},\frac{d-1}{2}\right) +
    \int_{0}^{\infty }dr r^{d-1}\int_{0}^{\pi /2}d\theta \sin
    ^{d-2}\theta \right. \nonumber
    \\
    && \times \left. \left[ \ln \left( -2r\sin ^2\theta I'_{r\cos \theta }(r\sin \theta )
    K'_{r\cos \theta }(r\sin \theta )\right) -2\sum_{l=1}^{M_1}
    \frac{V_{2l}(\cos \theta )}{(1+r^2)^l}\right] \right\} .
\end{eqnarray}
Under the condition $M_1>d/2-1$ the integral in this formula is
convergent. The numerical results for the finite part of the total
vacuum energy are given in Table \ref{tab:neumen}. In the
physically most important case $d=3$, the total Casimir energy for
the Neumann scalar induced by a single plate is positive. For the
total Casimir energy measured by an uniformly accelerated observer
one obtains
\begin{equation}\label{E1Ng}
E_{1Ng}=\frac{0.00292 g}{\xi _1^2},
\end{equation}
which is finite and independent on the normalization scale. As the
energy (\ref{E1Ng}) is positive, the corresponding vacuum forces
tend to decelerate the plate. Here the situation is opposite
compared to the case of the Dirichlet scalar.

\section{Casimir energy for two Neumann
plates}\label{sec:Neu2pl}

Consider the scalar vacuum in the region between two plates
located at $\xi =\xi _{1}$, $\xi =\xi _{2}$ and with the Neumann
boundary conditions on them:
\begin{equation}
\frac{\partial \varphi }{\partial \xi }|_{\xi =\xi
_{1}}=\frac{\partial \varphi }{\partial \xi }|_{\xi =\xi _{2}}=0.
\label{Nbound2pl}
\end{equation}
For the function $\phi (\xi )$ in expression (\ref{wavesracture})
one has
\begin{equation}\label{Niom}
\phi (\xi )=N_{i\omega }(k\xi ,k\xi _{2})\equiv I_{i\omega
}^{\prime }(k\xi _{2})K_{i\omega }(k\xi )-I_{i\omega }(k\xi
)K_{i\omega }^{\prime }(k\xi _{2}).
\end{equation}
From the boundary condition on the plate $\xi =\xi _1$ we obtain
that the corresponding eigenfrequencies are solutions to the
equation
\begin{equation}
N'_{i\omega }(k\xi _{1},k\xi _{2})\equiv I_{i\omega }^{\prime
}(k\xi _{2})K_{i\omega }^{\prime }(k\xi _{1})-I_{i\omega }^{\prime
}(k\xi _{1})K_{i\omega }^{\prime }(k\xi _{2})=0.  \label{Nmodes}
\end{equation}
We denote the positive roots to this equation by $\omega =\omega
_{Nn}(k\xi _{1},k\xi _{2})$, and will assume that they are
arranged in the ascending order $\omega _{Nn}<\omega _{Nn+1}$. For
the vacuum energy in the region $\xi _{1}\leq \xi \leq \xi _{2}$
per unit surface of the plates on has
\begin{equation}
E_{N}=\frac{1}{2}\int \frac{d^{d-1}k}{(2\pi
)^{d-1}}\sum_{n=1}^{\infty }\omega _{Nn}(k\xi _{1},k\xi _{2}).
\label{Neumen}
\end{equation}
We introduce the partial zeta function as
\begin{equation}
\zeta _{N}(s,k\xi _{1},k\xi _{2})=\sum_{n=1}^{\infty }\omega
_{Nn}^{-s}(k\xi _{1},k\xi _{2}). \label{zetaN}
\end{equation}
In order to obtain the Casimir energy, one has to find the
analytic continuation of (\ref{zetaN}) to $s=-1$. Employing the
Cauchy's theorem, zeta function (\ref{zetaN}) may be written under
the form of contour integral on the complex plane,
\begin{equation}
\zeta _{N}(s,k\xi _{1},k\xi _{2})=\frac{1}{2\pi i}\int_{C}dz\,z^{-s}\frac{%
\partial }{\partial z}\ln N'_{iz}(k\xi _{1},k\xi _{2}),  \label{zetaN1}
\end{equation}
with the same contour $C$ as in Eq. (\ref{zetaD1}). The integral
on the right of Eq. (\ref{zetaN1}) can be presented in the form
\begin{eqnarray}
\zeta _{N}(s,k\xi _{1},k\xi _{2}) &=&\zeta _{1N}^{(R)}(s,k\xi _{1})+\frac{1}{%
2\pi i}\sum_{\alpha =+,-}\int_{C^{\alpha }}dz\,z^{-s}\frac{\partial }{%
\partial z}\ln I_{-\alpha iz}^{\prime }(k\xi _{2})  \nonumber \\
&&+\frac{1}{2\pi i}\sum_{\alpha =+,-}\int_{C^{\alpha }}dz\,z^{-s}\frac{%
\partial }{\partial z}\ln \left[ 1-\frac{I_{-\alpha iz}^{\prime }(k\xi
_{1})K_{iz}^{\prime }(k\xi _{2})}{I_{-\alpha iz}^{\prime }(k\xi
_{2})K_{iz}^{\prime }(k\xi _{1})}\right]   \nonumber \\
&&+\frac{1}{2\pi i}\int_{C_{\rho }}dz\,z^{-s}\frac{\partial
}{\partial z}\ln N'_{iz}(k\xi _{1},k\xi _{2}), \label{zetaN2}
\end{eqnarray}
where $C^{+}$ and $C^{-}$ are the upper and lower halves of the
contour $C$ except the parts coming from the small semicircle
$C_{\rho }$. The last integral over $C_{\rho }$ vanishes in the
limit $\rho \rightarrow 0$ for ${\mathrm{Re}}s<2$ and it can be
omitted in the following consideration. After parametrizing the
integrals over the imaginary axis we obtain the formula
\begin{eqnarray}
\zeta _{N}(s,k\xi _{1},k\xi _{2}) &=&\zeta _{1N}^{(R)}(s,k\xi _{1})+\frac{1}{%
\pi }\sin \frac{\pi s}{2}\int_{\rho }^{\infty }dz\,z^{-s}\frac{\partial }{%
\partial z}\ln I_{z}^{\prime }(k\xi _{2})  \nonumber \\
&&+\frac{1}{\pi }\sin \frac{\pi s}{2}\int_{\rho }^{\infty }dz\,z^{-s}\frac{%
\partial }{\partial z}\ln \left[ 1-\frac{I_{z}^{\prime }(k\xi
_{1})K_{z}^{\prime }(k\xi _{2})}{I_{z}^{\prime }(k\xi
_{2})K_{z}^{\prime }(k\xi _{1})}\right] . \label{zetaN3}
\end{eqnarray}
The last integral on the right of this formula is finite at $s=-1$
and vanishes in the limits $\xi _{1}\rightarrow 0$ and $\xi
_{2}\rightarrow \infty $. It follows from here that the second
term on the right corresponds to the zeta function $\zeta
_{1N}^{(L)}(s,x)$ for the region on the left of a single plate
located at $\xi =\xi _{2}$. The procedure for the analytic
continuation of this function we have considered in the previous
section.

\begin{figure}[tbph]
\begin{center}
\epsfig{figure=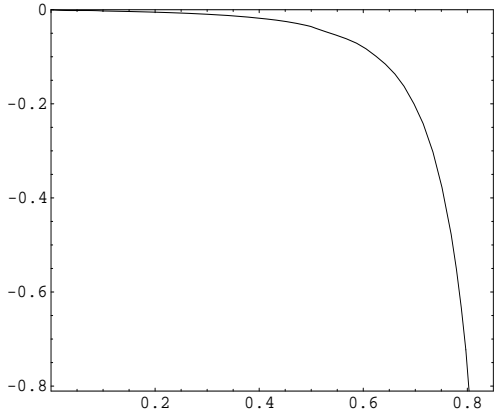,width=6.5cm,height=5.5cm}
\end{center}
\caption{Interference part of the Casimir energy in the region
between two Neumann plates, $\xi _2^{d-1}\Delta E_D(\xi _1,\xi
_2)$, as a function on the ratio $\xi _1/\xi _2$ for $d=3$.}
\label{fig3intn}
\end{figure}

Taking into account (\ref{zetaN2}), for the Neumann vacuum energy
in the region  $\xi _1\leq \xi \leq \xi _2$ one finds
\begin{equation}\label{EN2pl}
    E_N=E_{1N}^{(R)}(\xi _1)+E_{1N}^{(L)}(\xi _2)+\Delta E_N(\xi _1,\xi
    _2),
\end{equation}
with the interference term
\begin{equation}\label{ENint}
\Delta E_N(\xi _1,\xi _2)=\frac{B_d}{\pi } \int_{0}^{\infty }dk \,
k^{d-2} \int_{0}^{\infty }dz\, \ln \left[ 1-\frac{I'_z(k \xi
_1)K'_z(k \xi _2)}{I'_z(k \xi _2)K'_z(k \xi _1)}\right] ,
\end{equation}
where we have integrated in parts the $z$-integral. From the
inequality $N'_z(k\xi _1,k\xi _2)<0$ for $\xi _1<\xi _2$ it
follows that $\Delta E_N(\xi _1,\xi _2)<0$. In Fig. \ref{fig3intn}
we have presented the dependence of the interference part of the
Casimir energy (\ref{ENint}) on the ratio $\xi _1/\xi _2$ for
$d=3$.
As seen from Figures \ref{fig2intd} and \ref{fig3intn}, the
interference parts of the Casimir energies for Dirichlet and
Neumann scalars are numerically close to each other. This is a
consequence of that the subintegrands  in formulas (\ref{EDint})
and (\ref{ENint}) are numerically close. This can be also seen
analytically by using the inequalities for the Bessel modified
functions given in Ref. \cite{Avag02}. Interference term
(\ref{ENint}) is finite for all $0<\xi _1<\xi _2$, and diverges in
the limit $\xi _1\to \xi _2$. In this limit the main contribution
comes from the large values of $z$. Introducing a new integration
variable $x=k/z$ and using the uniform asymptotic expansions for
the Bessel modified functions, for the interference part in the
leading order one obtains the same result as in the Dirichlet
case, (\ref{DelEDlim1}). In particular, in the limit $\xi _1, \xi
_2 \to \infty$ with fixed $\xi _2-\xi _1$, for the vacuum energy
measured by an uniformly accelerated observer with the proper
acceleration $g\to 0$ and $g\xi _1\to 1$, one recovers the
standard result for the Casimir plates on the Minkowski bulk.

Adding to the energy in the region between the plates, given by
Eq. (\ref{EN2pl}), the energies coming from the regions $\xi \leq
\xi _1$ and $\xi \geq \xi _2$, we obtain the total Casimir energy,
$E_{D}^{{\mathrm{(tot)}}}$, for two-plates geometry:
\begin{equation}\label{EN2pltot}
E_N^{{\mathrm{(tot)}}}=E_{1N}(\xi _1)+E_{1N}(\xi _2)+\Delta
E_N(\xi _1,\xi _2),
\end{equation}
where the interference part is given by formula (\ref{ENint}). For
$d=3 $ Neumann scalar the first two terms on the right of Eq.
(\ref{EN2pltot}), corresponding to the single plates
contributions, are positive and the third (interference) term is
negative. For sufficiently close $\xi _1$ and $\xi _2$ the last
term dominates and the total Casimir energy is negative. For the
large separations between the plates the contributions from the
single plate parts ar dominant and the total energy is positive.
Hence, for some intermediate values of the separation the total
Casimir energy for the Neumann scalar vanishes.

As in the Dirichlet case, it can be seen that the energy
(\ref{ENint}) is related to the interaction forces between the
plates by the relations
\begin{subequations} \label{pN1int}
\begin{eqnarray}
\frac{\partial }{\partial \ln \xi _1}\Delta E_N(\xi _1,\xi _2)&=&
p_{N{\mathrm{(int)}}}^{(1)}(\xi _1,\xi _2)=\frac{B_d}{\pi \xi
_1^2} \int_{0}^{\infty }dk \, k^{d-2} \int_{0}^{\infty }dz\,
\frac{K'_z(k
\xi _2)(1+z^2/k^2\xi _1^2)}{K'_z(k \xi _1)N'_z(k\xi _1, k\xi _2)} \\
\frac{\partial }{\partial \ln \xi _2}\Delta E_N(\xi _1,\xi _2)&=&
-p_{N{\mathrm{(int)}}}^{(2)}(\xi _1,\xi _2)=-\frac{B_d}{\pi \xi
_2^2} \int_{0}^{\infty }dk \, k^{d-2} \int_{0}^{\infty }dz\,
\frac{I'_z(k \xi _1)(1+z^2/k^2\xi _2^2)}{I'_z(k \xi _2)N'_z(k\xi
_1, k\xi _2)} .
\end{eqnarray}
\end{subequations}
Here $p_{N{\mathrm{(int)}}}^{(j)}$, $j=1,2$ are the vacuum
effective pressures on the plate at $\xi =\xi _j$ induced by the
presence of the second plate. These quantities are investigated in
Ref. \cite{Avag02}.

\section{Casimir energy for the electromagnetic field} \label{sec:elmag}

We now turn to the case of the electromagnetic field. We will
assume that the plates are perfect conductors with the standard
boundary conditions of vanishing of the normal component of the
magnetic field and the tangential components of the electric
field, evaluated at the local inertial frame in which the
conductors are instantaneously at rest. As it has been shown in
Ref. \cite{Candelas} for $d=3$ and in Ref. \cite{Avag02} for an
arbitrary spatial dimension (on the decomposition of the electromagnetic field in the Rindler coordinates see also Ref. \cite{Gerl01}), the corresponding eigenfunctions for
the vector potential $A^{\mu }$ may be resolved into one transverse magnetic
(TM) mode and $d-2$ transverse electric (TE) (with respect to the
$\xi $ direction) modes $A^{\mu }_{\sigma \alpha }$, $\sigma =0,1,\ldots , d-2$, $\alpha =({\mathbf{k}},\omega)$:
\begin{eqnarray}
A^{\mu }_{1 \alpha }&=& \left( -\xi \frac{\partial }{\partial \xi },-i\frac{\omega }{\xi},0,\ldots ,0\right) \varphi _{0\alpha },\quad \sigma =1,\quad {\mathrm{TM\,\, mode}}, \label{TMA} \\
A^{\mu }_{\sigma \alpha } &=& \epsilon ^{\mu }_{\sigma }\varphi _{\sigma ,\alpha },\quad \sigma =0,2,\ldots ,d-2,\quad {\mathrm{TE\,\, modes}}, \label{TEA} .
\end{eqnarray}
Here the polarization vectors $\epsilon ^{\mu }_{\sigma }$ obey the following relations:
\begin{equation}
\epsilon ^{0}_{\sigma }=\epsilon ^{1}_{\sigma }=0, \quad \epsilon _{\sigma \mu} \epsilon ^{\mu }_{\sigma '}=-k^2\delta _{\sigma \sigma '},\quad \epsilon ^{\mu }_{\sigma }k_{\mu }=0.
\end{equation}
From the perfect conductor boundary
conditions one has the following conditions for the scalar fields $\varphi _{\sigma \alpha }$:
\begin{equation}
\varphi _{\sigma \alpha }\left| _{\xi =\xi _1}\right. =\varphi _{\sigma \alpha }\left| _{\xi =\xi _2}\right. =0, \quad \sigma =0,1,\ldots , d-2,
\label{TEbc}
\end{equation}
\begin{equation}
\frac{\partial \varphi _{1 \alpha }}{\partial \xi}\left| _{\xi =\xi _1}\right. =\frac{\partial \varphi _{1 \alpha }}{\partial \xi}\left| _{\xi =\xi _2}\right. =0 . \label{TMbc}
\end{equation}
As a result the TE/TM modes correspond to the
Dirichlet/Neumann scalars and the Casimir energy for
the electromagnetic field can be obtained from the scalar results
by the formula
\begin{equation}\label{Eelmag}
E_{{\mathrm{em}}} =(d-2)E_D+E_N .
\end{equation}
The same relation takes place for the energies
$E_{1{\mathrm{em}}}^{(R)}$, $E_{1{\mathrm{em}}}^{(L)}$, $\Delta
E_{{\mathrm{em}}}$, $E_{{\mathrm{em}}}^{{\mathrm{(tot)}}}$. In
particular, for the total electromagnetic Casimir energy in the
geometry of a single plate from numerical results given in Tables
\ref{tab:diren}, \ref{tab:neumen} one has $E_{{\mathrm{em}}}
=0.00160/\xi _1^2$ in $d=3$. Hence, in this case the
electromagnetic forces tend to decelerate the plate.

The total Casimir energy for two plates located at $\xi =\xi _1$
and $\xi = \xi _2$ can be written in the form
\begin{equation}\label{Eemtot}
E_{{\mathrm{em}}}^{{\mathrm{(tot)}}}=E_{1{\mathrm{em}}}(\xi
_1)+E_{1{\mathrm{em}}}(\xi _2) + \Delta E_{{\mathrm{em}}}(\xi
_1,\xi _2),
\end{equation}
where $\Delta E_{{\mathrm{em}}}(\xi _1,\xi _2)<0$. Note that the
quantity $\xi _1^{d-1}E_{{\mathrm{em}}}^{{\mathrm{(tot)}}}$ is a
function on the ratio $\xi _2/\xi _1$ only. For a given $\xi _1$
and for large values of this ratio the main contribution into
$E_{{\mathrm{em}}}^{{\mathrm{(tot)}}}$ comes from the first term
on the right of Eq. (\ref{Eemtot}). For $\xi _2/\xi _1-1\ll 1$ the
last term on the right of this formula dominates. From the
numerical results given above it follows that in $d=3 $ for a
given $\xi _1$ the energy $E_{{\mathrm{em}}}^{{\mathrm{(tot)}}}$
is positive for large values of $\xi _2/\xi _1$ and is negative
for $\xi _2/\xi _1-1\ll 1$. Hence, for some intermediate value of
$\xi _2/\xi _1$ this Casimir energy vanishes.

\section{Conclusion}\label{sec:conc}

In quantum field theory the different unitary inequivalent
representations of the commutation relations in general give rise
to different pictures with different physical implications, in
particular to different vacuum states. An interesting issue in the
investigations of the Casimir effect is the dependence of the
vacuum characteristics on the type of the vacuum. In this paper we
have investigated the Casimir energies generated by a single and
two parallel plates moving by uniform proper acceleration,
assuming that the fields are prepared in the Fulling-Rindler
vacuum state. The corresponding vacuum expectation values of the
energy--momentum tensor were investigated in Refs.
\cite{Candelas,Saha02} for the geometry of a single plate and in
Ref. \cite{Avag02} in the case of two plates. Due to the well
known surface divergencies in these expectation values, the total
Casimir energy cannot be evaluated by direct integration of the
vacuum energy density and needs an additional regularization. In
this paper as a regularization method we employ the zeta function
technique. We have considered the cases of scalar and
electromagnetic fields in an arbitrary number of the spacetime
dimensions.

For the scalar case both Dirichlet and Neumann boundary conditions
are investigated. In the case of a single plate geometry the right
Rindler wedge is divided into two regions, referred as RR and RL
regions. By using the Cauchiy's theorem on residues, we have
constructed an integral representations for the zeta functions in
both these regions, which are well suited for the analytic
continuation. Subtracting and adding to the integrands leading
terms of the corresponding uniform asymptotic expansions, we
present the corresponding functions $Z(s)$ as a sum of two parts.
The first one is convergent at $s=-1$ and can be evaluated
numerically. In the second, asymptotic part the pole contributions
are given explicitly in terms of beta function. As a consequence,
the Casimir energies for separate RR and RL regions contain pole
and finite contributions (see, for example, formulae (\ref{E1DR3})
and (\ref{E1DRp}) in the case of the Dirichlet scalar in the RR
region). The remained pole term is a characteristic feature for
the zeta function regularization method and has been found for
many other cases of boundary geometries. The coefficient for this
term is determined by the corresponding boundary coefficient in
the heat kernel asymptotic expansion. For an infinitely thin plate
taking RR and RL regions together, in odd spatial dimensions the
pole parts cancel and the Casimir energy for the whole Rindler
wedge is finite. Note that in this case the total Casimir energy
can be directly evaluated by making use formulae (\ref{E1Dtotal})
and (\ref{E1DNtotal}) for the Dirichlet and Neumann scalar fields,
respectively. The cancellation of the pole terms coming from
oppositely oriented faces of infinitely thin smooth boundaries
takes place in very many situations encountered in the literature.
It is a simple consequence of the fact that the second fundamental
forms are equal and opposite on the two faces of each boundary
and, consequently, the value of the corresponding coefficient in
the heat kernel expansion summed over the two faces of each
boundary vanishes \cite{Blau88}. In even dimensions there is no
such a cancellation. The numerical results for the separate pole
and finite contributions to the Casimir energy in RR and RL
regions are summarized in Table \ref{tab:diren} for the Dirichlet
case and in Table 2 for the Neumann boundary condition. For the
physically most important case $d=3$ the total Casimir energy is
negative for the Dirichlet scalar and positive for the Neumann
scalar. This means that the vacuum forces tend to accelerate the
Dirichlet plate and to decelerate the Neumann plate.

In the case of two parallel plates configuration we have derived
integral representations for both Dirichlet and Neumann zeta
functions. The corresponding Casimir energies are presented as a
sum of single plate parts and the interference term. The latter is
determined by formula (\ref{EDint}) for Dirichlet scalar and by
formula (\ref{ENint}) for Neumann scalar and is located in the
region between the plates. It is always negative and is related to
the corresponding interaction forces by the standard
thermodynamical relations (see Eqs. (\ref{pD1int}),
(\ref{pN1int})). For large values of the separation between the
plates, the total Casimir energy is dominated by the contributions
coming from the single plate parts. For a given $\xi _1$ and small
values of the separation the interference part is dominant and the
total Casimir energy is negative. For $d=3$ Dirichlet scalar this
is the case for all values of the separation. However, for $d=3$
Neumann scalar due to the positive single plate energies, the
total Casimir energy for two plates is negative for small
separations and positive for large separations. Consequently, this
energy vanishes for some intermediate value of the ratio $\xi
_2/\xi _1$. Letting $\xi _2\to \infty $ correspond to removing one of the plates and from the formulae for two plates we recover the Casimir energy for a single plate. The case $d=1$ is considered separately in Appendix
\ref{ap:d1}. In this dimension the Casimir energies for Dirichlet
and Neumann scalars are the same. They vanish for a single plate
(point) geometry and are negative for the two plates case. Note
that in $d=1$ the problem is conformally related to the
corresponding problem on background of the Minkowski spacetime.

In Sec. \ref{sec:elmag} the case of the electromagnetic field is
considered with the perfect conductor boundary conditions in the
local inertial frame in which the boundaries are instantaneously
at rest. The corresponding eigenmodes are superposition of TE and
TM modes with Dirichlet and Neumann boundary conditions,
respectively. The Casimir energies for the electromagnetic field
can be derived from the corresponding scalar results making use
formula (\ref{Eelmag}). In particular, the total electromagnetic
vacuum energy of a single plate in $d=3$ is positive and the
vacuum forces tend to decelerate the plate. For two plates
geometry the situation for the electromagnetic field is similar to
the Neumann scalar case. The total Casimir energy is negative for
small values of the plates separation (the interference part
dominates) and is positive for large separations (single plates
parts dominate).

In the main part of this paper we have considered the vacuum
energy corresponding to the dimensionless coordinate $\tau $ in
Eq. (\ref{metric}). The corresponding eigenfrequencies are also
dimensionless and there is no need to introduce an arbitrary mass
scale in the definitions of the related zeta functions. However,
this mass scale is necessary if we consider the vacuum energy
measured by an uniformly accelerated observer. The relation
between these energies is discussed in Appendix \ref{ap:propE},
where we have shown that they are connected by formula
(\ref{EgErel}). In this formula the logarithmic term with an
arbitrary mass scale $\mu $ has to be viewed as a remainder of the
renormalization process (for a discussion see
\cite{Bordag1,Blau88}). For infinitely thin boundaries in odd
dimensions, in calculations of the total vacuum energy, including
the parts from two sides of the boundary, these terms cancel and a
unique result emerges.

\section*{Acknowledgments}

The authors are grateful to Professor Edward Chubaryan and
Professor Roland Avagyan for general encouragement and
suggestions. AAS and AHY acknowledge the hospitality of the Abdus
Salam International Centre for Theoretical Physics, Trieste,
Italy. This work was supported in part by the Armenian Ministry of
Education and Science Grant No. 0887.

\appendix

\section{$d=1$ case}\label{ap:d1}

Let us consider the scalar vacuum in the region between two
boundaries located at $\xi = \xi_{1}$ and $\xi = \xi_{2}$. The
normalized eigenfunctions satisfying Dirichlet boundary conditions
and the corresponding eigenfrequencies are in the form
\cite{Avag02}
\begin{equation} \label{phiDd1}
\varphi_n^D=\frac{\exp ^{-i \omega _n \tau}}{\sqrt{\pi n}}\sin
\left( \omega _n \ln (\xi _2/\xi )\right) , \quad \omega
_n=\frac{\pi n}{\ln (\xi _2/\xi _1)}, \quad n=1,2,\ldots .
\end{equation}
For the corresponding vacuum energy one finds
\begin{equation}\label{EDd1}
    E_D=\frac{1}{2}\sum_{n=1}^{\infty }\omega _n=\frac{\pi \zeta _R(-1)}{2\ln (\xi _2/\xi
    _1)}=-\frac{\pi }{24\ln (\xi _2/\xi _1)},
\end{equation}
where $\zeta _R(x) $ is the Riemann zeta function. The
renormalized Casimir energies $E_{1D}^{(R)}$ and $E_{1D}^{(L)}$
for a single boundary are obtained from (\ref{EDd1}) in the limits
$\xi _1 \to 0$ and $\xi _2\to \infty $, respectively, and both
these quantities vanish.

For the case of the Neumann boundary conditions the eigenfunctions
have the form
\begin{equation} \label{phiNd1}
\varphi_n^N=\frac{\exp ^{-i \omega _n \tau}}{\sqrt{\pi n}}\cos
\left( \omega _n \ln (\xi _2/\xi )\right) , \quad  n=0,1,2,\ldots
\end{equation}
with the same eigenfrequencies as in Eq. (\ref{phiDd1}). The
corresponding Casimir energy is the same as in the Dirichlet case:
$E_N =E_D$.

\section{Relation to the energy measured by an uniformly accelerated observer}
\label{ap:propE}

In this appendix we consider the relation of the energies
evaluated above to the energy measured by an uniformly accelerated
observer. The frequency $\omega $ in Eq. (\ref{wavesracture})
corresponds to the dimensionless coordinate $\tau $ and, hence, is
dimensionless. Proper time $\tau _g$ and the frequency $\omega _g$
measured by an uniformly accelerated observer with the proper
acceleration $g$ and word line $(x^1)^2-t^2=g^{-2}$, are related
to $\tau $ and $\omega $ by formulae $\tau _g=\tau /g$, $\omega
_g=\omega g$ (the features of the measurements for time, frequency, and length relative to a Rindler frame as compared to a Minkowski frame are discussed in Ref. \cite{Gerl03}). The corresponding vacuum energy has the form
\begin{equation}\label{propE}
E_g=\frac{1}{2}\int\frac{d^{d-1}k}{(2\pi )^{d-1}}
\sum_{n=1}^{\infty }\omega _{gn},
\end{equation}
where $\omega _{gn}=\omega _{gn}(x)$ are the eigenfrequencies
measured by the uniformly accelerated observer. As above we
introduce the partial zeta function related to these
eigenfrequencies and the function $Z_g(s)$ by relations
\begin{equation}\label{Zgdsetag}
\zeta _g(s,x)=\sum_{n=1}^{\infty }(\omega _{gn}/\mu )^{-s},\quad
Z_g(s)=\int_{0}^{\infty } dx x^{d-2}\zeta _g(s,x).
\end{equation}
Note that we have introduced an arbitrary scale $\mu $ with mass
dimension, in order to keep the zeta function dimensionless for
all $s$. Now the Casimir energy can be written as
\begin{equation}\label{Egobs}
E_g=\frac{\mu B_d}{2\xi _1^{d-1}}Z_g(s)|_{s=-1}.
\end{equation}
By taking into account the relation between the eigenfrequencies
corresponding to $\tau $ and $\tau _g$ coordinates, one obtains
the relation between $Z_g(s) $ and the function $Z(s)$ considered
above: $Z_g(s)=(\mu /g)^s Z(s)$. Laurent-expanding over $s+1$ and
using the corresponding expansion for the function $Z(s)$ given
above, we obtain the following relation between the Casimir
energies:
\begin{equation}\label{EgErel}
E_{g}=g\left( E+\ln (\mu /g) \frac{B_d Z^{(as)}_{-1}}{\xi
_1^{d-1}}\right) ,
\end{equation}
where the energy $E $ corresponds to the dimensionless coordinate
$\tau$. In dependence of the boundary conditions  and the region
under consideration, here as a pair $(E,Z_{-1}^{(as)})$ we have to
take $(E^{(R)}_{1F},Z_{FK,-1}^{(as)})$,
$(E^{(L)}_{1F},Z_{FI,-1}^{(as)})$,
$(E_{1F},Z_{FK,-1}^{(as)}+Z_{FI,-1}^{(as)})$, where
$F=D,N,{\mathrm{em}}$. Note that in odd spatial dimensions for the
total Casimir energy $E_{1Fg}$ of a single plate the scale
dependent parts from the RR and RL regions cancel and we obtain a
scale independent energy.

\end{document}